\documentclass[%
superscriptaddress,
reprint,
nofootinbib,
amsmath,amssymb,amsfonts,
floatfix,
altaffilletter,
showkeys,
]{revtex4-2}
\setcitestyle{super}

\usepackage{microtype}
\usepackage{color}
\usepackage{graphicx}
\usepackage{caption}
\usepackage{subcaption}
\usepackage{booktabs} % for professional tables
\usepackage{multirow}
\usepackage{tabularx}
\usepackage{amsmath}
\usepackage{amssymb}
\usepackage{esvect} %nicer vectors with \vv{} instead of \vv{}
\usepackage{mathtools}
\usepackage{bm}
\usepackage[hidelinks]{hyperref}
\usepackage{pbox}
\usepackage{threeparttable}
\usepackage{xr}

\usepackage{pifont}

\newcommand{\nn}{SpookyNet}

\usepackage{xr}
\makeatletter
\newcommand*{\addFileDependency}[1]{
	\typeout{(#1)}
	\@addtofilelist{#1}
	\IfFileExists{#1}{}{\typeout{No file #1.}}
}
\makeatother

\newcommand*{\myexternaldocument}[1]{
	\externaldocument[S]{#1}
	\addFileDependency{#1.tex}
	\addFileDependency{#1.aux}
}
\myexternaldocument{supplement}

\makeatletter\@input{xxsupplement.tex}\makeatother
\begin{document}

\title{\nn{}: Learning Force Fields with \\Electronic Degrees of Freedom and Nonlocal Effects}
\author{Oliver T. Unke}
\email{oliver.unke@googlemail.com}
\affiliation{Machine Learning Group, Technische Universit\"at Berlin, 10587 Berlin, Germany}
\affiliation{DFG Cluster of Excellence ``Unifying Systems in Catalysis'' (UniSysCat), Technische Universit\"at Berlin, 10623 Berlin, Germany}

\author{Stefan Chmiela}
\affiliation{Machine Learning Group, Technische Universit\"at Berlin, 10587 Berlin, Germany}

\author{Michael Gastegger}
\affiliation{Machine Learning Group, Technische Universit\"at Berlin, 10587 Berlin, Germany}
\affiliation{DFG Cluster of Excellence ``Unifying Systems in Catalysis'' (UniSysCat), Technische Universit\"at Berlin, 10623 Berlin, Germany}

\author{Kristof T. Sch\"utt}
\affiliation{Machine Learning Group, Technische Universit\"at Berlin, 10587 Berlin, Germany}

\author{Huziel E.\ Sauceda}
\affiliation{Machine Learning Group, Technische Universit\"at Berlin, 10587 Berlin, Germany}
\affiliation{
	BASLEARN, BASF-TU joint Lab, Technische Universit\"at Berlin, 10587 Berlin, Germany
}

\author{Klaus-Robert M\"uller}
\email{klaus-robert.mueller@tu-berlin.de}
\affiliation{Machine Learning Group, Technische Universit\"at Berlin, 10587 Berlin, Germany}
\affiliation{Department of Artificial Intelligence, Korea University, Anam-dong, Seongbuk-gu, Seoul 02841, Korea}
\affiliation{Max Planck Institute for Informatics, Stuhlsatzenhausweg, 66123 Saarbr\"ucken, Germany}
\affiliation{BIFOLD -- Berlin Institute for the Foundations of Learning and Data, Berlin, Germany}
\affiliation{Google Research, Brain team, Berlin, Germany}

\begin{abstract}
Machine-learned force fields (ML-FFs) combine the accuracy of \textit{ab initio} methods with the efficiency of conventional force fields. However, current ML-FFs typically ignore electronic degrees of freedom, such as the total charge or spin state, and assume chemical locality, which is problematic when molecules have inconsistent electronic states, or when nonlocal effects play a significant role. This work introduces \nn{}, a deep neural network for constructing ML-FFs with explicit treatment of electronic degrees of freedom and quantum nonlocality. Chemically meaningful inductive biases and analytical corrections built into the network architecture allow it to properly model physical limits.
\nn{} improves upon the current state-of-the-art (or achieves similar performance) on popular quantum chemistry data sets. Notably, it is able to generalize across chemical and conformational space and can leverage the learned chemical insights, e.g.\ by predicting unknown spin states, thus helping to close a further important remaining gap for today's machine learning models in quantum chemistry.

\end{abstract}

\keywords{\nn{}, machine learning,  neural network, force field, potential energy surface, nonlocal, delocalized interactions, quantum nature, spooky, electronic degrees of freedom}

\maketitle

\section*{Introduction}
\label{sec:introduction}
\begin{figure*}
	\includegraphics[width=\textwidth]{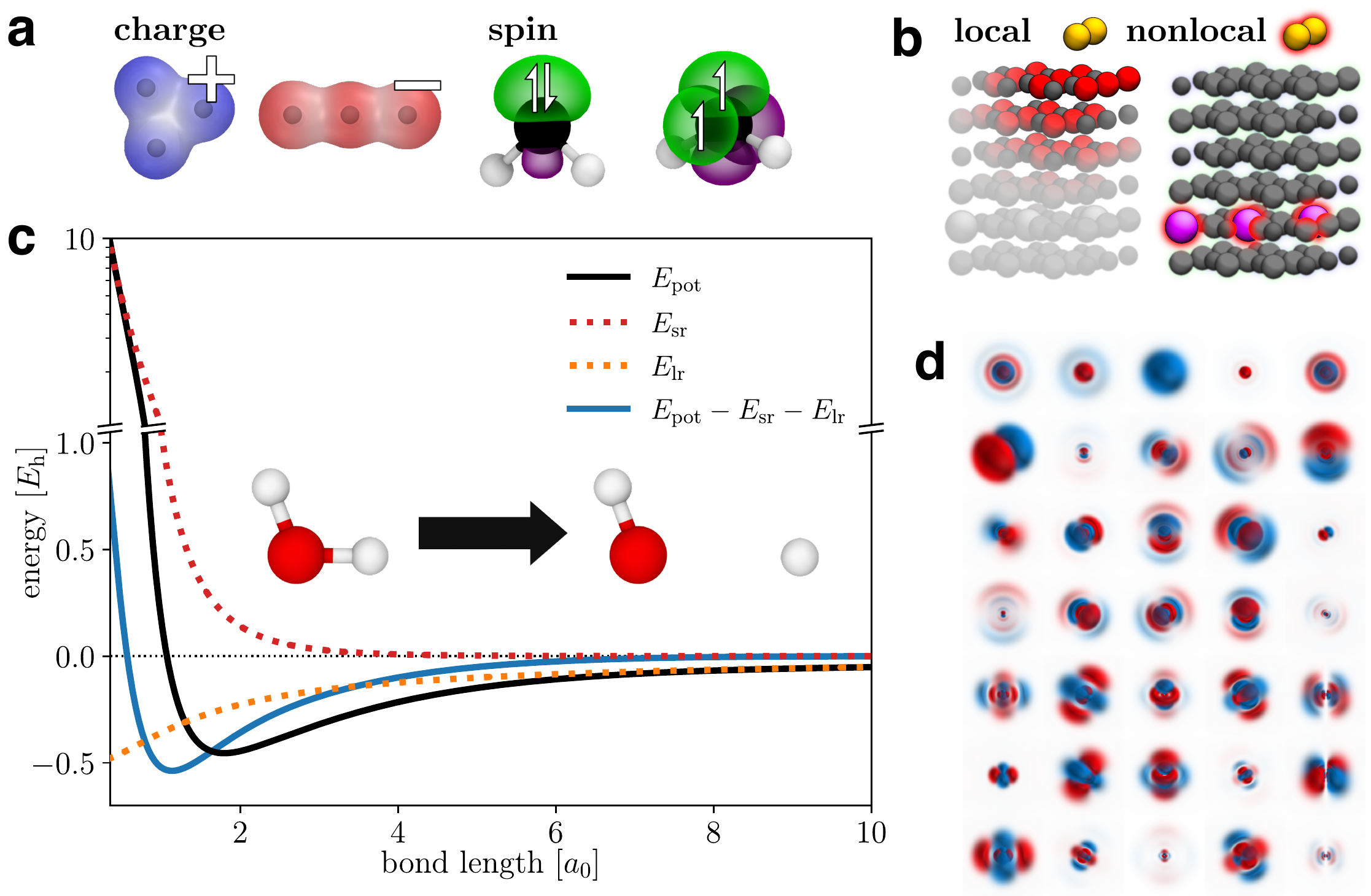}
	\caption{Main features of \nn{} and problems addressed in this work. (\textbf{a}) Optimized geometries of Ag$_3^+$/Ag$_3^-$ (left) and singlet/triplet CH$_2$ (right). Without information about the electronic state (charge/spin), machine learning models are unable to distinguish between the different structures. (\textbf{b}) Au$_2$ dimer on a MgO(001) surface doped with Al atoms (Au: yellow, Mg: grey, O: red, Al: pink). The presence of Al atoms in the crystal influences the electronic structure and affects Au$_2$ binding to the surface, an effect which cannot be adequately described by only local interactions. (\textbf{c}) Potential energy $E_{\rm pot}$ (solid black) for O--H bond dissociation in water. %(computed at the HF/6-31G level of theory).
	The asymptotic behavior of $E_{\rm pot}$ for very small and very large bond lengths can be well-approximated by analytical short-ranged $E_{\rm sr}$ (dotted red) and long-ranged $E_{\rm lr}$ (dotted orange) energy contributions, which follow known physical laws. When they are subtracted from $E_{\rm pot}$, the remaining energy (solid blue) covers a smaller range of values and decays to zero quicker, which simplifies the learning problem. (\textbf{d}) Visualization of a random selection of  learned interaction functions for \nn{} trained on the QM7-X\cite{hoja2021qm7} dataset. Note that they closely resemble atomic orbitals, demonstrating \nn{}'s ability to extract chemical insight from data.}
	\label{fig:overview}
\end{figure*}

Molecular dynamics (MD) simulations of chemical systems allow to gain insights on many intricate phenomena, such as reactions or the folding of proteins.\cite{warshel2002molecular} To perform MD simulations, knowledge of the forces acting on individual atoms at every time step of the simulation is required.\cite{karplus2002molecular} The most accurate way of deriving these forces is by (approximately) solving the Schr\"odinger equation~(SE), which describes the physical laws governing chemical systems.\cite{dirac1929quantum} Unfortunately,
the computational cost of accurate \textit{ab initio} approaches\cite{gordon2005theory} makes them impractical when many atoms are studied, or the simulation involves thousands (or even millions) of time steps. For this reason, it is common practice to use force fields (FFs) -- analytical expressions for the potential energy of a chemical system, from which forces are obtained by derivation -- instead of solving the SE.\cite{unke2020high}
The remaining difficulty is to find an appropriate functional form that gives forces at the required accuracy.

Recently, machine learning (ML) methods have gained increasing popularity for addressing this task.\cite{bartok2010gaussian,rupp2012fast,bartok2017machine,gastegger2018wacsf,schutt2019unifying,unke2020machine,von2020exploring,noe2020machinemolsim,glielmo2021unsupervised,keith2021combining} They allow to automatically learn the relation between chemical structure and forces from \emph{ab initio} reference data. The accuracy of such ML-FFs is limited by the quality of the data used to train them and their computational efficiency is comparable to conventional FFs.\cite{unke2020machine,sauceda2020molecular}

One of the first methods for constructing ML-FFs for high-dimensional systems was introduced by \citeauthor{behler2007generalized} for studying the properties of bulk silicon.\cite{behler2007generalized} The idea is to encode the local (within a certain cutoff radius) chemical environment of each atom in a descriptor, e.g.\ using symmetry functions,\cite{behler2011atom} which is used as input to an artificial neural network\cite{mcculloch1943logical} predicting atomic energies. The total potential energy of the system is modeled as the sum of the individual contributions, and forces are obtained by derivation with respect to atom positions. Alternatively it is also possible to directly predict the total energy (or forces) without relying on a partitioning into atomic contributions.\cite{unke2017toolkit,chmiela2017machine,chmiela2019sgdml} However, an atomic energy decomposition makes predictions extensive and the learned model applicable to systems of different size. Many other ML-FFs follow this design principle, but rely on different descriptors\cite{smith2017ani,zhang2018deep,unke2018reactive} or use other ML methods,\cite{bartok2010gaussian,bartok2017machine,christensen2020fchl} such as kernel machines,\cite{vapnik1995,cortes1995support,muller2001introduction,scholkopf2002learning,braun2008relevant,rupp2012fast,hansen2013assessment} for the prediction. An alternative to manually designed descriptors is to use the raw atomic numbers and Cartesian coordinates as input instead. Then, suitable atomic representations can be learned from (and adapted to) the reference data automatically. This is usually achieved by ``passing messages'' between atoms to iteratively build increasingly sophisticated descriptors in a deep neural network architecture. After the introduction of the deep tensor neural network (DTNN),\cite{schutt2017quantum} such message-passing neural networks (MPNNs)\cite{gilmer2017neural} became highly popular and the original architecture has since been refined by many related approaches.\cite{schutt2018schnet,lubbers2018hierarchical,unke2019physnet}

However, atomic numbers and Cartesian coordinates (or descriptors derived from them) do not provide an unambiguous description of chemical systems.\cite{friesner2005ab} They only account for the nuclear degrees of freedom, but contain no information about electronic structure, such as the total charge or spin state. This is of no concern when all systems of interest have a consistent electronic state (e.g.\ they are all neutral singlet structures), but leads to an ill-defined learning problem otherwise (Fig.~\ref{fig:overview}a). Further, most ML-FFs assume that atomic properties are dominated by their local chemical environment.\cite{unke2020machine} While this approximation is valid in many cases, it still neglects that quantum systems are inherently nonlocal in nature, a quality which Einstein famously referred to as ``\emph{spooky actions at a distance}''.\cite{born2005born} For example, electrons can be delocalized over a chemical system and charge or spin density may instantaneously redistribute to specific atoms based on distant structural changes (Fig.~\ref{fig:overview}b).\cite{noodleman1995orbital,dreuw2003long,duda2006resonant,bellec2010nonlocal,bostrom2018charge}

ML-FFs have only recently begun to address these issues. For example, the charge equilibration neural network technique~(CENT)\cite{ghasemi2015interatomic} was developed to construct interatomic potentials for ionic systems. In CENT, a neural network predicts atomic electronegativities (instead of energy contributions), from which partial charges are derived via a charge equilibration scheme\cite{rappe1991charge,wilmer2012extended,cheng2014charge} that minimizes the electrostatic energy of the system and models nonlocal charge transfer. Then, the total energy is obtained by an analytical expression involving the partial charges. Since they are constrained to conserve the total charge, different charge states of the same chemical system can be treated by a single model. The recently proposed fourth-generation Behler-Parinello neural network (4G-BPNN)\cite{ko2021fourth} expands on this idea using two separate neural networks: The first one predicts atomic electronegativities, from which partial charges are derived using the same method as in CENT. The second neural network predicts atomic energy contributions, receiving the partial charges as additional inputs, which contain implicit information about the total charge. The charge equilibration scheme used in CENT and 4G-BPNNs involves the solution of a system of linear equations, which formally scales cubically with the number of atoms, although iterative solvers can be used to reduce the complexity.\cite{ko2021fourth} Unfortunately, only different total charges, but not spin states, can be distinguished with this approach. In contrast, neural spin equilibration (NSE),\cite{zubatyuk2020teaching} a recently proposed modification to the AIMNet model,\cite{zubatyuk2019accurate} distinguishes between $\alpha$ and $\beta$-spin charges, allowing it to also treat different spin states. In the NSE method, a neural network predicts initial (spin) charges from descriptors that depend only on atomic numbers and coordinates. The discrepancy between predictions and true (spin) charges is then used to update the descriptors and the procedure is repeated until convergence.

The present work introduces \nn{}, a deep MPNN which takes atomic numbers, Cartesian coordinates, the number of electrons, and the spin state as direct inputs. It does not rely on equilibration schemes, which often involves the costly solution of a linear system, to encode the electronic state. Our end-to-end learning approach is shared by many recent ML methods that aim to solve the Schr\"odinger equation\cite{pfau2020ab,hermann2020deep,scherbela2021solving} and mirrors the inputs that are also used in \textit{ab initio} calculations. To model local interactions between atoms, early MPNNs relied on purely distance-based messages,\cite{schutt2017quantum,schutt2018schnet,unke2019physnet} whereas later architectures such as DimeNet\cite{klicpera2020directional} proposed to include angular information in the feature updates. However, explicitly computing angles between all neighbors of an atom scales quadratically with the number of neighbors. To achieve linear scaling, \nn{} encodes angular information implicitly via the use of novel basis functions based on Bernstein polynomials\cite{bernstein1912demo} and spherical harmonics. Spherical harmonics are also used in neural network architectures for predicting rotationally equivariant quantities, such as tensor field networks,\cite{thomas2018tensor} Cormorant,\cite{anderson2019cormorant} PaiNN,\cite{schutt2021equivariant} or NequIP.\cite{batzner2021se} However, since only scalar quantities (energies) need to be predicted for constructing ML-FFs, \nn{} projects rotationally equivariant features to invariant representations for computational efficiency. Many methods for constructing descriptors of atomic environments use similar approaches to derive rotationally invariant quantities from spherical harmonics.\cite{bartok2013representing,thompson2015spectral,unke2018reactive}
In addition, \nn{} allows to model quantum nonlocality and electron delocalization explicitly by introducing a nonlocal interaction between atoms, which is independent of their distance. Its energy predictions are augmented with physically-motivated corrections to improve the description of long-ranged electrostatic and dispersion interactions and short-ranged repulsion between nuclei, which simplify the learning problem and guarantee correct asymptotic behaviour (Fig.~\ref{fig:overview}c). Further inductive biases in \nn{}'s architecture encourage learning of atomic representations which capture similarities between different elements and interaction functions which resemble atomic orbitals, allowing it to efficiently extract meaningful  chemical insights from data (Fig.~\ref{fig:overview}d).

\section*{Results}
\label{sec:results}

\begin{figure*}
    \centering
    \includegraphics[width=\textwidth]{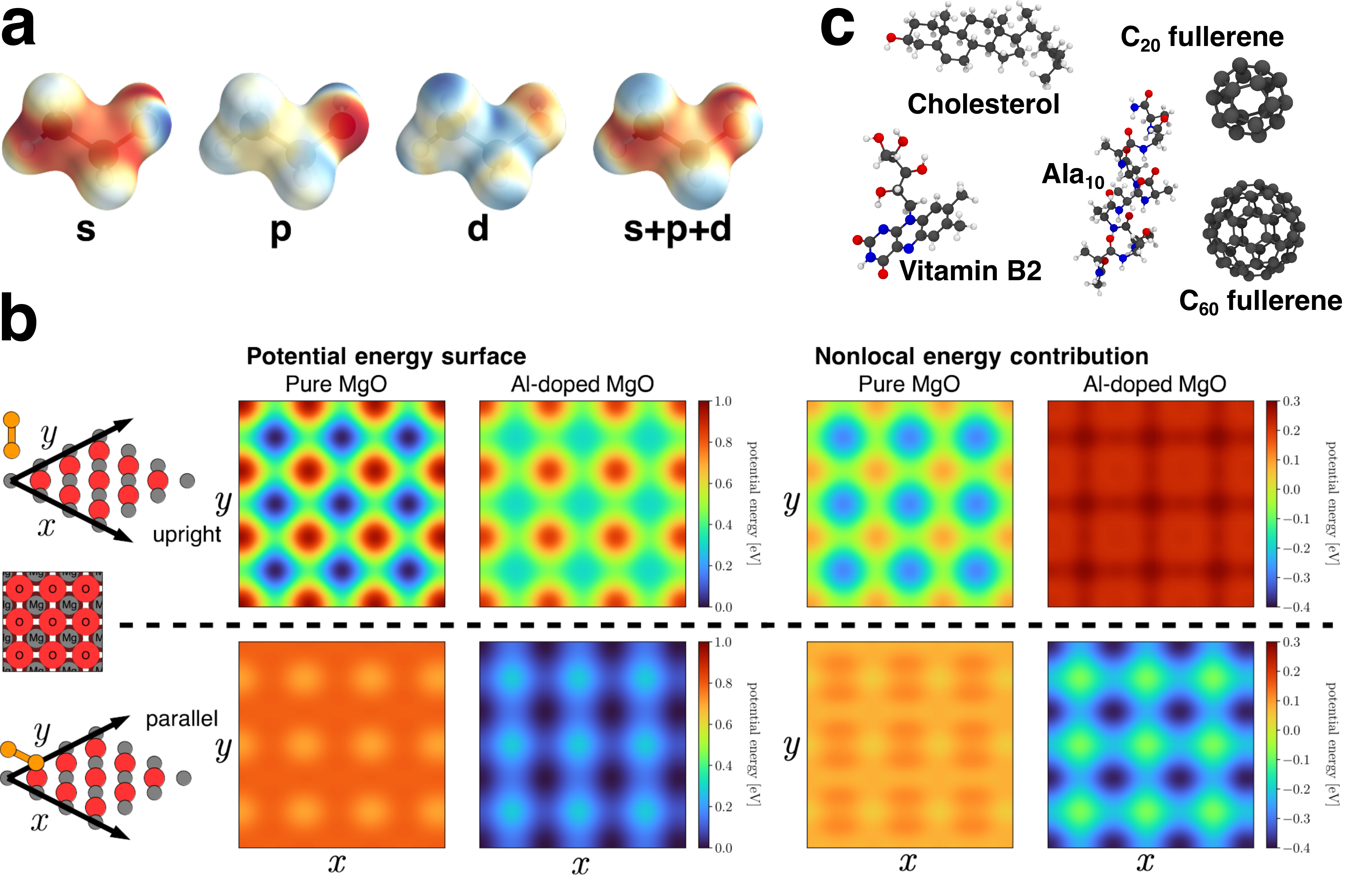}
    \caption{Examples of chemical insights extracted by \nn{}. (\textbf{a}) Visualization of the learned local chemical potential for ethanol (see methods). The individual contributions of s-, p-, and d-orbital-like interactions are shown (red: low energy, blue: high energy). (\textbf{b}) Potential energy surface scans obtained by moving an Au$_2$ dimer over an (Al-doped) MgO surface in different (upright/parallel) configurations (the positions of Mg and O atoms are shown for orientation). \nn{} learns to distinguish between local and nonlocal contributions to the potential energy, allowing it to model changes of the potential energy surface when the crystal is doped with Al atoms far from the surface. (\textbf{c}) A model trained on small organic molecules learns general chemical principles that transfer to much larger structures outside the chemical space covered by the training data. Here, optimized geometries obtained from \nn{} trained on the QM7-X database are shown.}
    \label{fig:insights}
\end{figure*}

\subsection*{\nn{} architecture}

\begin{figure*}
	\includegraphics[width=\textwidth]{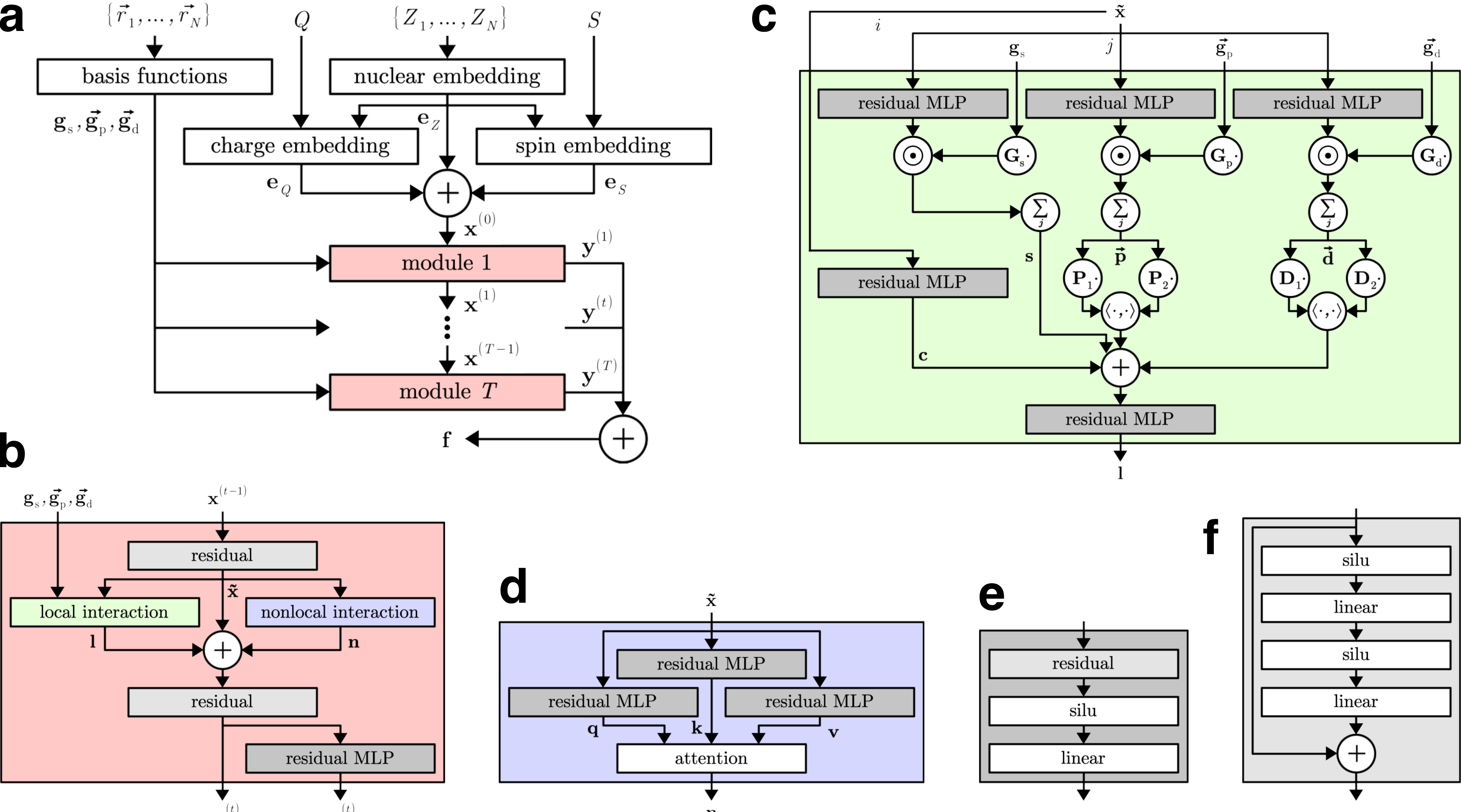}
	\caption{Schematic depiction of the \nn{} architecture with color-coded view of individual components. (\textbf{a}) Architecture overview, for details on the nuclear and electronic (charge/spin) embeddings and basis functions, refer to Eqs.~\ref{eq:nuclear_embedding},~\ref{eq:electronic_embedding}~and~\ref{eq:basis_functions}, respectively. (\textbf{b}) Interaction module, see Eq.~\ref{eq:interaction_module}. (\textbf{c}) Local interaction block, see Eq.~\ref{eq:local_interaction}. (\textbf{d}) Nonlocal interaction block, see Eq.~\ref{eq:nonlocal_interaction}. (\textbf{e}) Residual multilayer perceptron (MLP), see Eq.~\ref{eq:resmlp}. (\textbf{f}) Pre-activation residual block, see Eq.~\ref{eq:residual}.}
	\label{fig:network_overview}
\end{figure*}

\nn{} takes sets of atomic numbers $\{Z_1,\dots,Z_N \mid Z_i\in\mathbb{N}\}$ and Cartesian coordinates $\{\vv{r}_1,\dots,\vv{r}_N\mid \vv{r}_i\in\mathbb{R}^3 \}$, which describe the element types and positions of~$N$ atoms, as input. Information about the electronic wave function, which is necessary for an unambiguous description of a chemical system, is provided via two additional inputs: The total charge $Q\in\mathbb{Z}$ encodes the number of electrons (given by $Q + \sum_i Z_i$), whereas the total angular momentum is encoded as the number of unpaired electrons $S\in\mathbb{N}_0$. For example, a singlet state is indicated by $S=0$, a doublet state by $S=1$, and so on.
The nuclear charges $Z$, total charge $Q$ and spin state $S$ are transformed to $F$-dimensional embeddings and combined to form initial atomic feature representations
\begin{equation}
\mathbf{x}^{(0)} = \mathbf{e}_Z + \mathbf{e}_Q + \mathbf{e}_S \,.
\label{eq:initial_atomic_features}
\end{equation}
Here, the nuclear embeddings $\mathbf{e}_Z$ contain information about the ground state electron configuration of each element and the electronic embeddings $\mathbf{e}_Q$ and $\mathbf{e}_S$ contain delocalized information about the total charge and spin state, respectively.
A chain of $T$ interaction modules iteratively refines these representations through local and nonlocal interactions
\begin{equation}
\begin{aligned}
\mathbf{x}_i^{(t)} = \mathbf{x}_i^{(t-1)} &+ \mathrm{local}^{(t)}\left(\{\mathbf{x}_j^{(t-1)}, \vv{r}_{ij}\}_{j\in\mathcal{N}(i)}\right)\\
 &+\mathrm{nonlocal}^{(t)}\left(\{\mathbf{x}^{(t-1)}\}\right)\,,
\end{aligned}
\label{eq:high_level_interaction}
\end{equation}
where $\mathcal{N}(i)$ contains all atom indices within a cutoff distance $r_{\rm cut}$ of atom~$i$ and $\vv{r}_{ij} = \vv{r}_{j}-\vv{r}_{i}$ is the relative position of atom~$j$ with respect to atom~$i$. The local interaction functions resemble s, p, and d atomic orbitals (see Fig.~\ref{fig:overview}d) and the model learns to encode different distance and angular information about the local environment of each atom with the different interaction functions (see Fig.~\ref{fig:insights}a). The nonlocal interactions on the other hand model the delocalized electrons.
The representations $\mathbf{x}^{(t)}$ at each stage are further refined through learned functions $\mathcal{F}_t$ according to $\mathbf{y}_i^{(t)}=\mathcal{F}_t(\mathbf{x}^{(t)})$ and summed to the atomic descriptors
\begin{equation}
\mathbf{f} = \sum_{t=1}^{T}\mathbf{y}_i^{(t)}\,,
\label{eq:atomic_descriptor}
\end{equation}
from which atomic energy contributions $E_i$ are predicted with linear regression. The total potential energy is given by
\begin{equation}
E_{\rm pot} = \sum_{i=1}^{N}E_i + E_{\rm rep} + E_{\rm ele} + E_{\rm vdw}\,,
\label{eq:potential_energy}
\end{equation}
where $E_{\rm rep}$, $E_{\rm ele}$, and $E_{\rm vdw}$ are empirical corrections, which augment the energy prediction with physical knowledge about nuclear repulsion, electrostatics, and dispersion interactions. Energy-conserving forces $\vv{F}_i = -\partial E_{\rm pot}/\partial \vv{r}_i$ can be obtained via automatic differentiation. A schematic depiction of the \nn{} architecture is given in Fig.~\ref{fig:network_overview}.

\subsection*{Electronic states}
\label{sec:electronic_states}
\begin{figure}
	\includegraphics[width=\columnwidth]{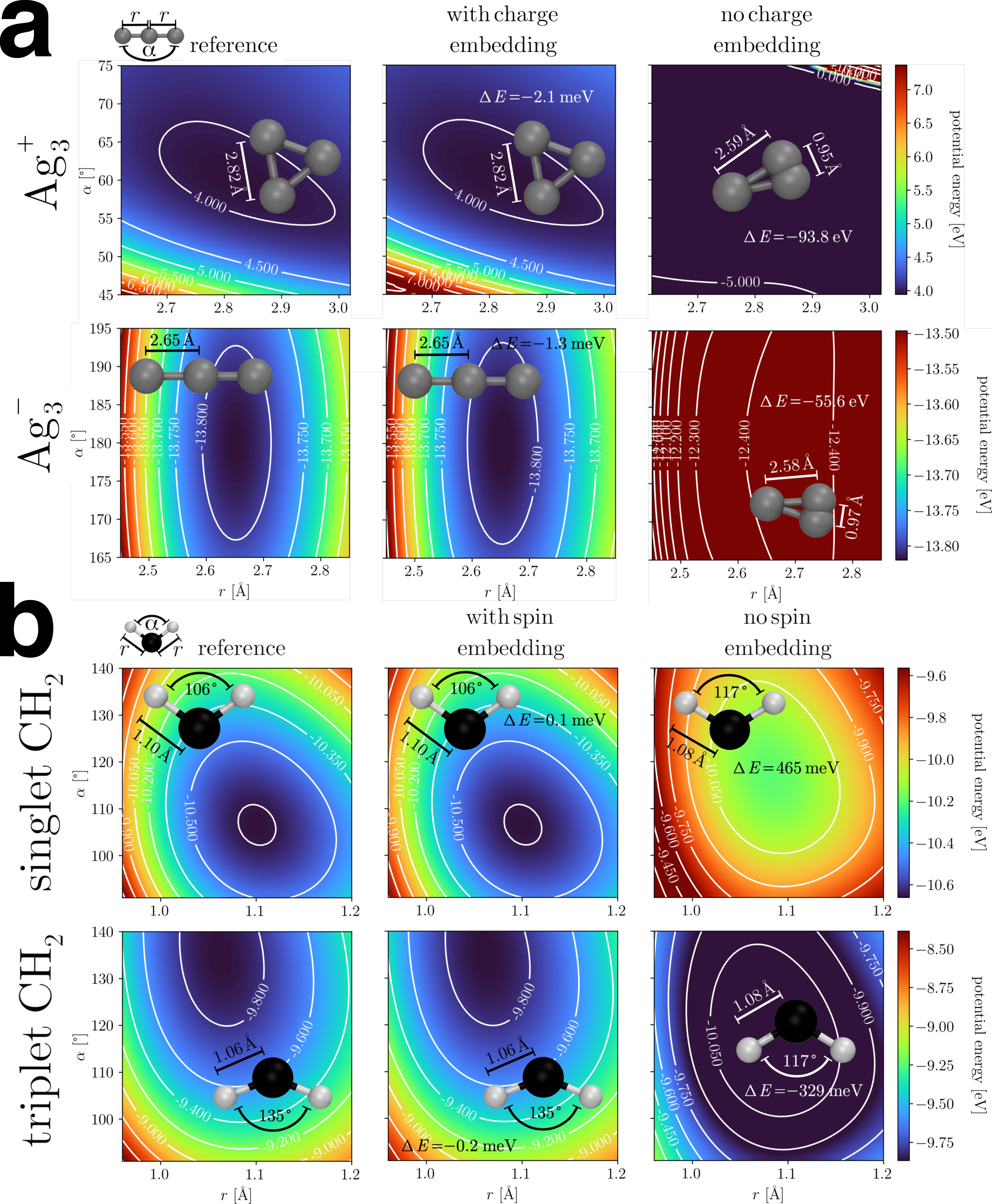}
	\caption{Potential energy surfaces of (\textbf{a}) Ag$_3^+$ and Ag$_3^-$ and (\textbf{b}) singlet and triplet CH$_2$ predicted by \nn{} with (middle column) and without (right column) charge/spin embedding (the reference ground truth is shown in the left column). Minimum energy structures and prediction errors ($\Delta E$) for the minimum energy are also shown.}
	\label{fig:electronic_states}
\end{figure}

Most existing ML-FFs can only model structures with a consistent electronic state, e.g.\ neutral singlets. An exception are systems for which the electronic state can be inferred via structural cues, e.g.\ in the case of protonation/deprotonation.\cite{unke2019physnet} In most cases, however, this is not possible, and ML-FFs that do not model electronic degrees of freedom are unable to capture the relevant physics. Here, this problem is solved by explicitly accounting for different electronic states (see Eq.~\ref{eq:initial_atomic_features}). To illustrate their effects on potential energy surfaces, two exemplary systems are considered: Ag$_3^+$/Ag$_3^-$ and singlet/triplet~CH$_2$, which can only be distinguished by their charge, respectively their spin state. \nn{} is able to faithfully reproduce the reference potential energy surface for all systems. When the charge/spin embeddings $\mathbf{e}_Q$/$\mathbf{e}_S$ (Eq.~\ref{eq:initial_atomic_features}) are removed, however,
the model becomes unable to represent the true potential energy surface and its predictions are qualitatively different from the reference (see Fig.~\ref{fig:electronic_states}). As a consequence, wrong global minima are predicted when performing geometry optimizations with a model trained without the charge/spin embeddings, whereas they are virtually indistinguishable from the reference when the embeddings are used. Interestingly, even without a charge embedding, \nn{} can predict different potential energy surfaces for Ag$_3^+$/Ag$_3^-$. This is because explicit point charge electrostatics are used in the energy prediction (see Eq.~\ref{eq:potential_energy}) and the atomic partial charges are constrained such that the total molecular charge is conserved. However, such implicit information is insufficient to distinguish both charge states adequately. In the case of singlet/triplet~CH$_2$, there is no such implicit information and both systems appear identical to a model without electronic embeddings, i.e.\ it predicts the same energy surface for both systems, which neither reproduces the true singlet nor triplet reference.

Models with electronic embeddings even generalize to unknown electronic states. As an example, the QMspin database\cite{qmspin} is considered. It consists of $\sim$13k carbene structures with at most nine non-hydrogen atoms (C, N, O, F), which were optimized either in a singlet or triplet state. For each of these, both singlet and triplet energies computed at the MRCISD+Q-F12/cc-pVDZ-F12 level of theory are reported, giving a total of $\sim$26k energy-structure pairs in the database (see Ref.~\citenum{schwilk2020large} for more details). For the lack of other models evaluated on this data set, \nn{} is compared to itself without electronic embeddings. This baseline model only reaches a mean absolute prediction error (MAE) of 444.6~meV for unknown spin states. As expected, the performance is drastically improved when the electronic embeddings are included, allowing \nn{} to reach an MAE of 68.0~meV. Both models were trained on 20k points, used 1k samples as validation set, and were tested on the remaining data.
An analysis of the local chemical potential (see methods) reveals that a model with electronic embeddings learns a feature-rich internal representation of molecules, which significantly differs between singlet and tripled states (see Fig.~\ref{Sfig:different_spin_gummybears}). In contrast, the local chemical potential learned by a model without electronic embeddings is almost uniform and (necessarily) identical between both states, suggesting that the electronic embeddings are crucial to extract the relevant chemistry from the data.

%\begin{table}
%	\begin{tabular}{l r r}
%		\toprule
%		& \bf MAE & \bf RMSE \\
%		\midrule
%		\bf with spin embedding & 68.0 & 132.3\\
%		\bf no spin embedding & 444.6 & 571.3 \\
%		\bottomrule
%	\end{tabular}
%	\caption{Mean absolute errors (MAEs) and root mean square errors (RMSEs) in meV for predicting unknown spin states in the QMSpin database\cite{qmspin}.}
%	\label{tab:qmspin_results}
%\end{table}

\subsection*{Nonlocal effects}
\label{sec:nonlocal_effects}
\begin{figure*}
	\includegraphics[width=\textwidth]{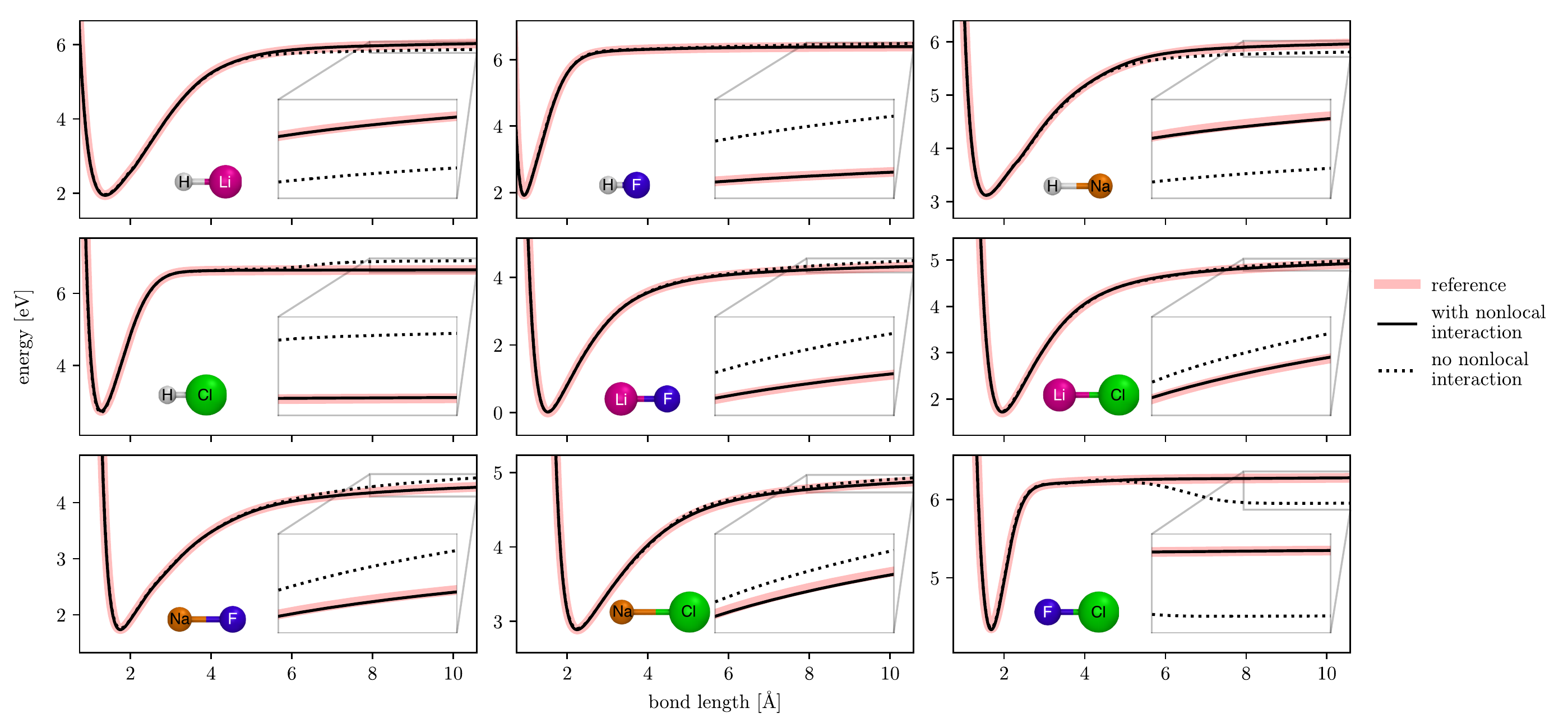}
	\caption{Dissociation curves of different diatomic molecules predicted by \nn{} with/without nonlocal interactions.}
	\label{fig:nonlocal}
\end{figure*}

For many chemical systems, local interactions are sufficient for an accurate description. However, there are also cases were a purely local picture breaks down. To demonstrate that nonlocal effects can play an important role even in simple systems, the dissociation curves of nine (neutral singlet) diatomic molecules made up of H, Li, F, Na, and Cl atoms are considered (Fig.~\ref{fig:nonlocal}). Once the bonding partners are separated by more than the chosen cutoff radius $r_{\rm cut}$, models that rely only on local information will always predict the same energy contributions for atoms of the same element (by construction). However, since electrons are free to distribute unevenly across atoms, even when they are separated (e.g.\ due to differences in their electronegativity), energy contributions should always depend on the presence of other atoms in the structure. Consequently, it is difficult for models without nonlocal interactions to predict the correct asymptotic behavior for all systems simultaneously. As such, when the nonlocal interactions are removed from interaction modules (Eq.~\ref{eq:high_level_interaction}), \nn{} predicts an unphysical ``step'' for large interatomic separations, even when a large cutoff is used for the local interactions. In contrast, the reference is reproduced faithfully when nonlocal interactions are enabled. Note that such artifacts -- occurring if nonlocal interactions are not modeled -- are problematic e.g.\ when simulating reactions. Simply increasing the cutoff is no adequate solution to this problem, since it just shifts the artifact to larger separations. %Similarly, the inclusion of long-range corrections is also insufficient to avoid problems in the asymptotic tails (all analytical corrections for predicting the potential energy are enabled for both models).
%In this particular example, some problems associated with the lack of nonlocal interactions can be partially compensated when training on just a few chemical systems. For example, when fitted on just one or two out of the nine systems, \nn{} is able to account for different asymptotes with element-specific energy bias terms. When trained on all nine dissociation curves, however, there are insufficient degrees of freedom in the bias parameters to account for all asymptotes in this way and the nonlocal effects must be modeled explicitly.

More complex nonlocal effects may arise for larger structures. For example, \citeauthor{ko2021fourth} recently introduced four data sets for systems exhibiting nonlocal charge transfer effects.\cite{ko2021fourth}
One of these systems
consists of a diatomic Au cluster deposited on the surface of a periodic MgO(001) crystal (Au$_2$\nobreakdash--MgO). In its minimum energy configuration, the Au$_2$ cluster ``stands upright'' on the surface on top of an O atom. When some of the Mg atoms (far from the surface) are replaced by Al (see Fig.~\ref{fig:overview}b), however, the Au$_2$ cluster prefers to ``lie parallel'' to the surface above two Mg atoms (the distance between the Au and Al atoms is above 10~\AA). In other words, the presence of Al dopant atoms nonlocally modifies the electronic structure at the surface in such a way that a different Au$_2$ configuration becomes more stable. This effect can be quantified by scanning the potential energy surface of Au$_2$\nobreakdash--MgO by moving the Au$_2$ cluster above the surface in different configurations (see Fig.~\ref{fig:insights}b). Upon introduction of dopant Al atoms, nonlocal energy contributions destabilize the upright configuration of Au$_2$, particularly strongly above the positions of oxygen atoms. In contrast, the parallel configuration is lowered in energy, most strongly above positions of Mg atoms.

When applied to the Au$_2$\nobreakdash--MgO system, \nn{} significantly improves upon the values reported for models without any treatment of nonlocal effects and also achieves lower prediction errors than 4G\nobreakdash-BPNNs,\cite{ko2021fourth} which model nonlocal charge transfer via charge equilibration (see Table~\ref{tab:behler_results}).
For completeness, values for the other three systems introduced in Ref.~\citenum{ko2021fourth} are also reported in Table~\ref{tab:behler_results}, even though they could be modeled without including nonlocal interactions (as long as charge embeddings are used). For details on the number of training/validation data used for each data set, refer to Ref.~\citenum{ko2021fourth} (all models use the same settings).

\begin{table*}
	\begin{tabular}{c l c c c c}
		\toprule
		& & \bf 2G-BPNN & \bf 3G-BPNN & \bf 4G-BPNN & \bf \nn{} \\
		\midrule
		\multirow{3}{*}{\bf C$_{10}$H$_2$/C$_{10}$H$_3^+$} & \it energy & 1.619 & 2.045 & 1.194 & \bf 0.364 \\
		& \it forces & 129.5 & 231.0 & 78.00 & \bf 5.802 \\
		& \it charges & --- & 20.08 & 6.577 & \bf 0.117\\
		\midrule
		\multirow{3}{*}{\bf Na$_{8/9}$Cl$_8^+$} & \it energy & 1.692 & 2.042 & 0.481 & \bf 0.135 \\
		& \it forces & 57.39 & 76.67 & 32.78 & \bf 1.052 \\
		& \it charges & --- & 20.80 & 15.83 & \bf 0.111\\
		\midrule
		\multirow{3}{*}{\bf Ag$_3^{+/-}$} & energy & 352.0 & 320.2 & 1.323 & \bf 0.220 \\
		& forces & 1803 & 1913 & 31.69 & \bf 26.64\\
		& charges & --- & 26.48 & 9.976 & \bf 0.459\\
		\midrule
		\multirow{3}{*}{\bf Au$_2$--MgO} &  \it energy & 2.287 & --- & 0.219 & \bf 0.107 \\
		& \it forces & 153.1 & --- & 66.0 & \bf 5.337 \\
		& \it charges & --- & --- & 5.698 & \bf 1.013\\
		\bottomrule
	\end{tabular}
	\caption{Root mean square errors (RMSEs) of energies (meV/atom), forces
		(meV~\AA$^{-1}$) and charges (me) for the datasets introduced in Ref.~\citenum{ko2021fourth}. The values for 2G-, 3G-, and 4G-BPNNs are taken from Ref.~\citenum{ko2021fourth}. Best results in bold.}
	\label{tab:behler_results}
\end{table*}

\subsection*{Generalization across chemical and conformational space}
\label{sec:generalization}
For more typical ML-FF construction tasks where nonlocal effects are negligible and all molecules have consistent electronic states, \nn{} improves upon the generalization capabilities of comparable ML-FFs. Here, the QM7-X database\cite{hoja2021qm7} is considered as a challenging benchmark. This database was generated starting from $\sim$7k molecular graphs with up to seven non-hydrogen atoms (C, N, O, S, Cl) drawn from the GDB13 chemical universe.\cite{blum2009970} Structural and constitutional (stereo)isomers were sampled and optimized for each graph, leading to $\sim$42k equilibrium structures. For each of these, an additional 100 non-equilibrium structures were generated by displacing atoms along linear combinations of normal modes corresponding to a temperature of 1500~K, which leads to $\sim$4.2M structures in total. For each of these, QM7-X contains data for 42 physicochemical properties (see Ref.~\citenum{hoja2021qm7} for a complete list). For constructing ML-FFs, however, the properties $E_{\rm tot}$ and $F_{\rm tot}$, which correspond to energies and forces computed at the PBE0+MBD\cite{adamo1999toward,tkatchenko2012accurate} level of theory, are the most relevant. %Within non-equilibrium conformations sampled around the same minimum structure, the energy varies between $\sim$3--20~eV and forces between $\sim$10--103~eV~\AA$^{-1}$, i.e.\ the range of energies and forces is almost one order of magnitude larger than in MD17.

Because of the variety of molecules and the strongly distorted conformers contained in the QM7-X data set, models need to be able to generalize across both chemical and conformational space to perform well. Here, two different settings are considered: In the more challenging task (unknown molecules/unknown conformations), a total of 10\,100 entries corresponding to all structures sampled for 25 out of the original $\sim$7k molecular graphs are reserved as test set and models are trained on the remainder of the data. In this setting, all structures in the test set correspond to unknown molecules, i.e.\ the model has to learn general principles of chemistry to perform well. %This setting is especially interesting for the purpose of constructing ML-FFs that can be applied to a large variety of different molecules.
As comparison, an easier task (known molecules/unknown conformations) is constructed by randomly choosing 10\,100 entries as test set, so it is very likely that the training set contains at least some structures for all molecules contained in QM7-X (only unknown conformations need to be predicted).  \nn{} achieves lower prediction errors than both SchNet\cite{schutt2018schnet} and PaiNN\cite{schutt2021equivariant} for both tasks and is only marginally worse when predicting completely unknown molecules, suggesting that it successfully generalizes across chemical space (see Table~\ref{tab:qm7x_results}).
\begin{table*}
	\begin{tabular}{c c c c  c}
		\toprule
		task & &
		\textbf{SchNet}\cite{schutt2018schnet} & \textbf{PaiNN}\cite{schutt2021equivariant} &
		\textbf{\nn{}} \\
		\midrule
		\textbf{known molecules/} &  \textit{energy} & 50.847 & 15.691 & \textbf{10.620} (0.403) \\
		\textbf{unknown conformations} &  \textit{forces} & 53.695 & 20.301 & \textbf{14.851} (0.430) \\
		\midrule
		\textbf{unknown molecules/} &  \textit{energy} & 51.275 & 17.594 & \textbf{13.151} (0.423)  \\
		\textbf{unknown conformations} & \textit{forces} & 62.770  & 24.161 & \textbf{17.326} (0.701) \\
		\bottomrule
	\end{tabular}
	\caption{Mean absolute errors for energy (meV) and force (meV~\AA$^{-1}$) predictions for the QM7-X\cite{hoja2021qm7} dataset. Results for \nn{} are averaged over four runs, the standard deviation between runs is given in brackets. Best results in bold.}
	\label{tab:qm7x_results}
\end{table*}
Interestingly, a \nn{} model trained on QM7-X also generalizes to significantly larger chemical structures: Even though it was trained on structures with at most seven non-hydrogen atoms, it can be used e.g.\ for geometry optimizations of molecules like vitamin B2, cholesterol, or deca-alanine (see Fig.~\ref{fig:insights}c). Remarkably, it even predicts the correct structures for fullerenes, although the QM7-X dataset contains no training data for any pure carbon structure.

Since the QM7-X dataset has only recently been published, the performance of \nn{} is also benchmarked on the well-established MD17 data set.\cite{chmiela2017machine}
MD17 consists of structures, energies, and forces collected from \textit{ab initio} MD simulations of small organic molecules at the PBE+TS\cite{perdew1996generalized,tkatchenko2009accurate} level of theory.  %Trajectories were calculated at a temperature of 500~K, a resolution of 0.5~fs, and contain between $\sim$10$^5$--10$^6$ conformations.
Prediction errors for several models published in the literature are summarized in Table~\ref{tab:md17_results} and compared to \nn{}, which reaches lower prediction errors or closely matches the performance of other models for all tested molecules.
\begin{table*}
	\begin{tabular}{c c c c c c c c}
		\toprule
		& & \bf sGDML\cite{chmiela2019sgdml} & \bf SchNet\cite{schutt2018schnet} & \bf PhysNet\cite{unke2019physnet} & \bf FCHL19\cite{christensen2020fchl} & \bf PaiNN\cite{schutt2021equivariant} & \bf \nn{} \\
		\midrule
		\multirow{2}{*}{\bf Aspirin} & \it energy & 0.19 & 0.37 & 0.230 & 0.182 & 0.159 & {\bf 0.151} (0.008) \\
		& \it forces & 0.68 & 1.35 & 0.605 & 0.478 & 0.371 & {\bf 0.258} (0.034) \\
		\midrule
		\multirow{2}{*}{\bf Ethanol} & \it energy & 0.07 & 0.08 & 0.059 & 0.054 & 0.063 & {\bf 0.052} (0.001)\\
		& \it forces & 0.33 & 0.39 & 0.160 & 0.136 & 0.230 & {\bf 0.094} (0.011) \\
		\midrule
		\multirow{2}{*}{\bf Malondialdehyde} & \it energy & 0.10 & 0.13 & 0.094 & 0.081 & 0.091 & {\bf 0.079} (0.002)\\
		& \it forces & 0.41 & 0.66 & 0.319 & 0.245 & 0.319 & {\bf 0.167} (0.015) \\
		\midrule
		\multirow{2}{*}{\bf Naphthalene} & \it energy & 0.12 & 0.16 & 0.142 & 0.117 & 0.117 & {\bf 0.116} (0.001)\\
		& \it forces & 0.11 & 0.58 & 0.310 & 0.151 & \bf 0.083 & 0.089 (0.018) \\
		\midrule
		\multirow{2}{*}{\bf Salicylic acid} & \it energy & 0.12 & 0.20 & 0.126 & \bf 0.114 & \bf 0.114 & {\bf 0.114} (0.004)\\
		& \it forces & 0.28 & 0.85 & 0.337 & 0.221 & 0.209 & {\bf 0.180} (0.040) \\
		\midrule
		\multirow{2}{*}{\bf Toluene} & \it energy & 0.10 & 0.12 & 0.100 & 0.098 & 0.097 & {\bf 0.094} (0.001) \\
		& \it forces & 0.14 & 0.57 & 0.191 & 0.203 & 0.102 & {\bf 0.087} (0.014)\\
		\midrule
		\multirow{2}{*}{\bf Uracil} & \it energy & 0.11 & 0.14 & 0.108 & \bf 0.104 & \bf 0.104 & 0.105 (0.001)\\
		& \it forces & 0.24 & 0.56 & 0.218 & \bf 0.105 & 0.140 & 0.119 (0.021)\\
		\bottomrule
	\end{tabular}
	\caption{Mean absolute errors for energy (kcal~mol$^{-1}$) and force  (kcal~mol$^{-1}$~\AA$^{-1}$) predictions for the MD17 benchmark. Results for \nn{} are averaged over ten random splits, the standard deviation between runs is given in brackets. All models are trained on 1000 data points (separate models are used for each molecule), best results in bold.}
	\label{tab:md17_results}
\end{table*}\\

\section*{Discussion}
\label{sec:discussion}

The present work introduced \nn{}, an MPNN for constructing ML-FFs, which models electronic degrees of freedom and nonlocal interactions using attention.\cite{vaswani2017attention,choromanski2020rethinking}
\nn{} includes physically motivated inductive biases that facilitate the extraction of chemical insight from data. For example, element embeddings in \nn{} include the ground state electronic configuration, which encourages alchemically meaningful representations. An analytical short-range correction based on the Ziegler-Biersack-Littmark stopping potential\cite{ziegler1985stopping} improves the description of nuclear repulsion, whereas long-range contributions to the potential energy are modeled with point charge electrostatics and an empirical dispersion correction, following previous works.\cite{artrith2011high,morawietz2012neural,morawietz2013density,uteva2017interpolation,yao2018tensormol,unke2019physnet}
%Formally, evaluating such long-range contributions scales quadratically with the number of atoms, but methods to achieve linearithmic scaling for evaluating long-ranged interactions in conventional force fields can also be applied here.\cite{darden1993particle,fennell2006ewald}
These empirical augmentations allow \nn{} to extrapolate beyond the data it was trained on based on physical knowledge from data.

\nn{} can predict different potential energy surfaces for the same molecule in different electronic states and is able to model nonlocal changes to the properties of materials such as MgO upon introduction of dopant atoms. Further, it successfully generalizes to structures well outside the chemical and conformational space covered by its training data and improves upon existing models in different quantum chemical benchmarks. The interaction functions learned by \nn{} resemble atomic orbitals (see Fig.~\ref{fig:overview}d), demonstrating that it represents molecular systems in a chemically intuitive manner (see also Fig.~\ref{fig:insights}a). Obtaining such an understanding of how ML models,\cite{samek2019explainable} here \nn{},  solve a prediction problem is crucially important in the sciences as a low test set error\cite{hansen2013assessment} alone can not rule out that a model may overfit or for example capitalize on various artifacts in data\cite{samek2021explaining} or show ``Clever Hans'' effects.\cite{lapuschkin2019unmasking}

So far, most ML-FFs rely on nuclear charges and atomic coordinates as their only inputs and are thus unable to distinguish chemical systems with different electronic states. Further, they often rely on purely local information and break down when nonlocal effects cannot be neglected. The novel additions to MPNN architectures introduced in this work solve both of these issues, extending the applicability of ML-FFs to a much wider range of chemical systems than was previously possible and allow to model properties of quantum systems that have been neglected in many existing ML-FFs.

Remaining challenges in the construction of ML-FFs pertain to their successful application to large and heterogenuous condensed phase systems, such as proteins in aqueous solution. This is a demanding task, among others, due to the difficulty of performing \textit{ab initio} calculations for such large systems, which is necessary to generate appropriate reference data. Although models trained on small molecules may generalize well to larger structures, it is not understood how to guarantee that all relevant regions of the potential energy surface, visited e.g.\ during a dynamics simulation, are well described. We
conjecture that the inclusion of physically motivated inductive biases, which is a crucial ingredient in the \nn{} architecture, may serve as a general design principle to improve the next generation of ML-FFs and tackle such problems.

\section*{Methods}
\label{sec:methods}

\subsection*{Details on the neural network architecture}
\label{sec:neural_network_architecture}

%\paragraph*{Notation} Scalar values are denoted by either lowercase~$a$ or uppercase~$A$ symbols. Vectors in abstract feature spaces are indicated by bold lowercase symbols~$\mathbf{v}$ to distinguish them from vectors with directional information (in physical space), which are written with an accented arrow~$\vv{v}$. The notation $\vv{\mathbf{v}}$ indicates a feature vector with an additional directional dimension, in contrast to matrices, which are written with bold uppercase symbols~$\mathbf{M}$. Norms $\lVert\cdot\rVert$ and scalar products $\langle \cdot,\cdot \rangle$ are always computed across directional dimensions and Hadamard (element-wise) multiplication~$\odot$ implies scalar multiplication for entries with directional information. Subscript indices~$i$ are used to indicate that a quantity is affiliated with atom~$i$, but are omitted whenever a distinction between different atoms is not necessary for clarity.\\

In the following, basic neural network building blocks and components of the \nn{} architecture are described in detail (see Fig.~\ref{fig:network_overview} for a schematic depiction). A standard building block of most neural networks are linear layers, which take input features $\mathbf{x}\in\mathbb{R}^{n_{\rm in}}$ and transform them according to
\begin{equation}
\mathrm{linear}(\mathbf{x}) = \mathbf{W}\mathbf{x} + \mathbf{b}\,,
\label{eq:linear}
\end{equation}
where $\mathbf{W}\in\mathbb{R}^{n_{\rm out}\times n_{\rm in}}$ and $\mathbf{b}\in\mathbb{R}^{n_{\rm out}}$ are learnable weights and biases, and $n_{\rm in}$ and $n_{\rm out}$ are the dimensions of the input and output feature space, respectively (in this work, $n_{\rm in}=n_{\rm out}$ unless otherwise specified).
Since Eq.~\ref{eq:linear} can only describe linear transformations, an activation function is required to learn nonlinear mappings between feature spaces.
Here, a generalized SiLU (\underline{Si}gmoid \underline{L}inear \underline{U}nit) activation function\cite{hendrycks2016gaussian,elfwing2018sigmoid} (also known as ``$\mathrm{swish}$''\cite{ramachandran2017searching}) given by
\begin{equation}
\mathrm{silu}(x) = \frac{\alpha x}{1+e^{-\beta x}}
\label{eq:activation}
\end{equation}
is used. Depending on the values of $\alpha$ and $\beta$, Eq.~\ref{eq:activation} smoothly interpolates between a linear function and the popular ReLU (\underline{Re}ctified \underline{L}inear \underline{U}nit) activation\cite{nair2010rectified} (see Fig.~\ref{Sfig:activation}).
Instead of choosing arbitrary fixed values, $\alpha$ and $\beta$ are learnable parameters in this work. Whenever the notation $\mathrm{silu}(\mathbf{x})$ is used, Eq.~\ref{eq:activation} is applied to the vector~$\mathbf{x}$ entry-wise and separate $\alpha$ and $\beta$ parameters are used for each entry. Note that a smooth activation function is necessary for predicting potential energies, because the presence of kinks would introduce discontinuities in the atomic forces.

%A combination of just two linear layers with (most) nonlinear activation functions is sufficient to approximate arbitrary functional relations between inputs and outputs, provided that a sufficiently large feature space is used in the \textit{hidden} layer (intermediate layer between inputs and outputs).\cite{gybenko1989approximation,hornik1991approximation} However, \emph{deep} neural networks with many hidden layers stacked on top of each other often perform better in practice and were shown to be more parameter-efficient for approximating some functional relations.\cite{eldan2016power,cohen2016expressive,telgarsky2016benefits,lu2017expressive}
In theory, increasing the number of layers should never decrease the performance of a neural network, since in principle, superfluous layers could always learn the identity mapping. In practice, however, deeper neural networks become increasingly difficult to train due to the vanishing gradients problem,\cite{glorot2010understanding} which often degrades performance when too many layers are used.
To combat this issue, it is common practice to introduce ``shortcuts'' into the architecture that skip one or several layers,\cite{srivastava2015highway} creating a \emph{residual block}.\cite{he2016deep} By inverting the order of linear layers and activation functions, it is even possible to train neural networks with several hundreds of layers.\cite{he2016identity} These ``pre-activation'' residual blocks transform input features $\mathbf{x}$ according to
\begin{equation}
\mathrm{residual}(\mathbf{x}) = \mathbf{x} + \mathrm{linear}_2(\mathrm{silu}_2(\mathrm{linear}_1(\mathrm{silu}_1(\mathbf{x}))))\,.
\label{eq:residual}
\end{equation}

Throughout the \nn{} architecture, small feedforward neural networks consisting of a residual block, followed by an activation and a linear output layer, are used as learnable feature transformations. For conciseness, such residual multilayer perceptrons (MLPs) are written as
\begin{equation}
\mathrm{resmlp}(\mathbf{x}) = \mathrm{linear}(\mathrm{silu}(\mathrm{residual}(\mathbf{x}))) \,.
\label{eq:resmlp}
\end{equation}

The inputs to \nn{} are transformed to initial atomic features (Eq.~\ref{eq:initial_atomic_features}) via embeddings. A nuclear embedding is used to map atomic numbers $Z\in\mathbb{N}$ to vectors $\mathbf{e}_Z\in\mathbb{R}^F$ given by
\begin{equation}
\mathbf{e}_Z =
\mathbf{M}\mathbf{d}_Z + \tilde{\mathbf{e}}_Z \,.
\label{eq:nuclear_embedding}
\end{equation}
Here, $\mathbf{M} \in \mathbb{R}^{F\times 20}$ is a parameter matrix that projects constant element descriptors $\mathbf{d}_Z\in \mathbb{R}^{20}$ to an $F$-dimensional feature space and $\tilde{\mathbf{e}}_Z \in \mathbb{R}^{F}$ are element-specific bias parameters. The descriptors $\mathbf{d}_Z$ encode information about the ground state electronic configuration of each element (see Table~\ref{Stab:species_descriptor} for details). Note that the term $\tilde{\mathbf{e}}_Z$ by itself allows to learn arbitrary embeddings for different elements, but including $\mathbf{M}\mathbf{d}_Z$ provides an inductive bias to learn representations that capture similarities between different elements, i.e.\ contain alchemical knowledge.

Electronic embeddings are used to map the total charge $Q\in\mathbb{Z}$ and number of unpaired electrons $S\in\mathbb{N}_0$ to vectors $\mathbf{e}_Q,\mathbf{e}_S\in\mathbb{R}^F$, which delocalize this information over all atoms via a mechanism similar to attention.\cite{vaswani2017attention} The mapping is given by
\begin{equation}
\begin{aligned}
\mathbf{q} &= \mathrm{linear}(\mathbf{e}_{Z}) ,\,
\mathbf{k} = \begin{cases}
\tilde{\mathbf{k}}_{+} & \Psi \geq 0\\
\tilde{\mathbf{k}}_{-} & \Psi < 0
\end{cases} ,\,
\mathbf{v} = \begin{cases}
\tilde{\mathbf{v}}_{+} & \Psi \geq 0\\
\tilde{\mathbf{v}}_{-} & \Psi < 0
\end{cases} ,\\
a_i &=  \frac{\Psi\ln\left(1+\exp\left(\mathbf{q}_i^{\mathsf{T}}\mathbf{k}/\sqrt{F}\right)\right)}{\sum_{j=1}^{N}\ln\left(1+\exp\left(\mathbf{q}_j^{\mathsf{T}}\mathbf{k}/\sqrt{F}\right)\right) } ,\,
\mathbf{e}_{\Psi} = \mathrm{resmlp}(a\mathbf{v}) \,,
\end{aligned}
\label{eq:electronic_embedding}
\end{equation}
where $\tilde{\mathbf{k}},\tilde{\mathbf{v}} \in \mathbb{R}^F$ are parameters and $\Psi=Q$ for charge embeddings, or $\Psi=S$ for spin embeddings (independent parameters are used for each type of electronic embedding). Separate parameters indicated by subscripts $+/-$ are used for positive and negative total charge inputs $Q$ (since $S$ is always positive or zero, only the $+$\nobreakdash-parameters are used for spin embeddings). Here, all bias terms in the $\mathrm{resmlp}$ transformation (Eq.~\ref{eq:resmlp}) are removed, such that when $a\mathbf{v} = \mathbf{0}$, the electronic embedding $\mathbf{e}_{\Psi} = \mathbf{0}$ as well. Note that $\sum_{i}a_i = \Psi$, i.e.\ the electronic information is distributed across atoms with weights proportional to the scaled dot product $\mathbf{q}_i^{\mathsf{T}}\mathbf{k}/\sqrt{F}$.

The initial atomic representations $\mathbf{x}^{(0)}$ (Eq.~\ref{eq:initial_atomic_features}) are refined iteratively by a chain of $T$ interaction modules according to
\begin{equation}
\begin{aligned}
\tilde{\mathbf{x}} &= \mathrm{residual}_1(\mathbf{x}^{(t-1)})\,,\\
\mathbf{x}^{(t)} &= \mathrm{residual}_2(\tilde{\mathbf{x}} + \mathbf{l} + \mathbf{n})\,,\\
\mathbf{y}^{(t)} &= \mathrm{resmlp}(\mathbf{x}^{(t)})\,.
\end{aligned}
\label{eq:interaction_module}
\end{equation}
Here, $\tilde{\mathbf{x}}\in\mathbb{R}^F$ are temporary atomic features and $\mathbf{l},\mathbf{n}\in \mathbb{R}^F$ represent interactions with other atoms. They are computed by local (Eq.~\ref{eq:local_interaction}) and nonlocal (Eq.~\ref{eq:nonlocal_interaction}) interaction blocks, respectively, which are described below.
Each module $t$ produces two outputs $\mathbf{x}^{(t)},\mathbf{y}^{(t)}\in \mathbb{R}^F$, where $\mathbf{x}^{(t)}$ is the input to the next module in the chain and all $\mathbf{y}^{(t)}$ outputs are accumulated to the final atomic descriptors $\mathbf{f}$ (Eq.~\ref{eq:atomic_descriptor}).

The features $\mathbf{l}$ in Eq.~\ref{eq:interaction_module} represent a local interaction of atoms within a cutoff radius $r_{\rm cut}\in\mathbb{R}$ and introduce information about the atom positions~$\vv{r}\in\mathbb{R}^3$. They are computed from the temporary features $\tilde{\mathbf{x}}$ (see Eq.~\ref{eq:interaction_module}) according to
\begin{equation}
\begin{aligned}
\mathbf{c} &= \mathrm{resmlp}_\mathrm{c}(\tilde{\mathbf{x}})\,,\\
\mathbf{s}_i &= \sum_{j\in\mathcal{N}(i)}   \mathrm{resmlp}_\mathrm{s}(\tilde{\mathbf{x}}_j) \odot \left(\mathbf{G}_\mathrm{s}\mathbf{g}_\mathrm{s}(\vv{r}_{ij})\right)\,,\\
\vv{\mathbf{p}}_i &= \sum_{j\in \mathcal{N}(i)}  \mathrm{resmlp}_\mathrm{p}(\tilde{\mathbf{x}}_j) \odot \left(\mathbf{G}_\mathrm{p}\vv{\mathbf{g}}_\mathrm{p}(\vv{r}_{ij})\right)\,,\\
\vv{\mathbf{d}}_i &= \sum_{j\in \mathcal{N}(i)}   \mathrm{resmlp}_\mathrm{d}(\tilde{\mathbf{x}}_j) \odot \left(\mathbf{G}_\mathrm{d}\vv{\mathbf{g}}_\mathrm{d}(\vv{r}_{ij})\right)\,,\\
\mathbf{l} &= \mathrm{resmlp}_\mathrm{l}\left(\mathbf{c} + \mathbf{s} + \langle\mathbf{P}_1\vv{\mathbf{p}},\mathbf{P}_2\vv{\mathbf{p}}\rangle + \langle\mathbf{D}_1\vv{\mathbf{d}},\mathbf{D}_2\vv{\mathbf{d}}\rangle\right)\,,
\end{aligned}
\label{eq:local_interaction}
\end{equation}
where, $\mathcal{N}(i)$ is the set of all indices $j\neq i$ for which $\lVert \vv{r}_{ij}\rVert < r_{\rm cut}$ (with $\vv{r}_{ij} = \vv{r}_{j}-\vv{r}_{i}$). The parameter matrices $\mathbf{G}_\mathrm{s},\mathbf{G}_\mathrm{p},\mathbf{G}_\mathrm{d}\in\mathbb{R}^{F\times K}$ are used to construct feature-wise interaction functions as linear combinations of basis functions $\mathbf{g}_\mathrm{s}\in\mathbb{R}^K$, $\vv{\mathbf{g}}_\mathrm{p}\in\mathbb{R}^{K\times3}$, and  $\vv{\mathbf{g}}_\mathrm{d}\in\mathbb{R}^{K\times5}$ (see Eq.~\ref{eq:basis_functions}), which have the same rotational symmetries as s-, p-, and d-orbitals. The features $\mathbf{s}\in\mathbb{R}^{F}$,  $\vv{\mathbf{p}}\in\mathbb{R}^{F\times3}$, and $\vv{\mathbf{d}}\in\mathbb{R}^{F\times5}$ encode the arrangement of neighboring atoms within the cutoff radius and $\mathbf{c}\in\mathbb{R}^{F}$ describes the central atom in each neighborhood. Here, $\mathbf{s}$ stores purely radial information, whereas $\vv{\mathbf{p}}$ and $\vv{\mathbf{d}}$ allow to resolve angular information in a computationally efficient manner (see Section~\ref{Ssec:atomic_descriptors} in the Supporting Information for details).  The parameter matrices $\mathbf{P}_1, \mathbf{P}_2, \mathbf{D}_1, \mathbf{D}_2 \in \mathbb{R}^{F\times F}$ are used to compute two independent linear projections for each of the rotationally equivariant features $\vv{\mathbf{p}}$ and $\vv{\mathbf{d}}$, from which rotationally invariant features are obtained via a scalar product $\langle\cdot,\cdot\rangle$. %Contrary to simply taking the norm over the directional dimension, this method of scalarization allows non-trivial mixing between different feature channels.
\begin{figure}
	\includegraphics[width=\columnwidth]{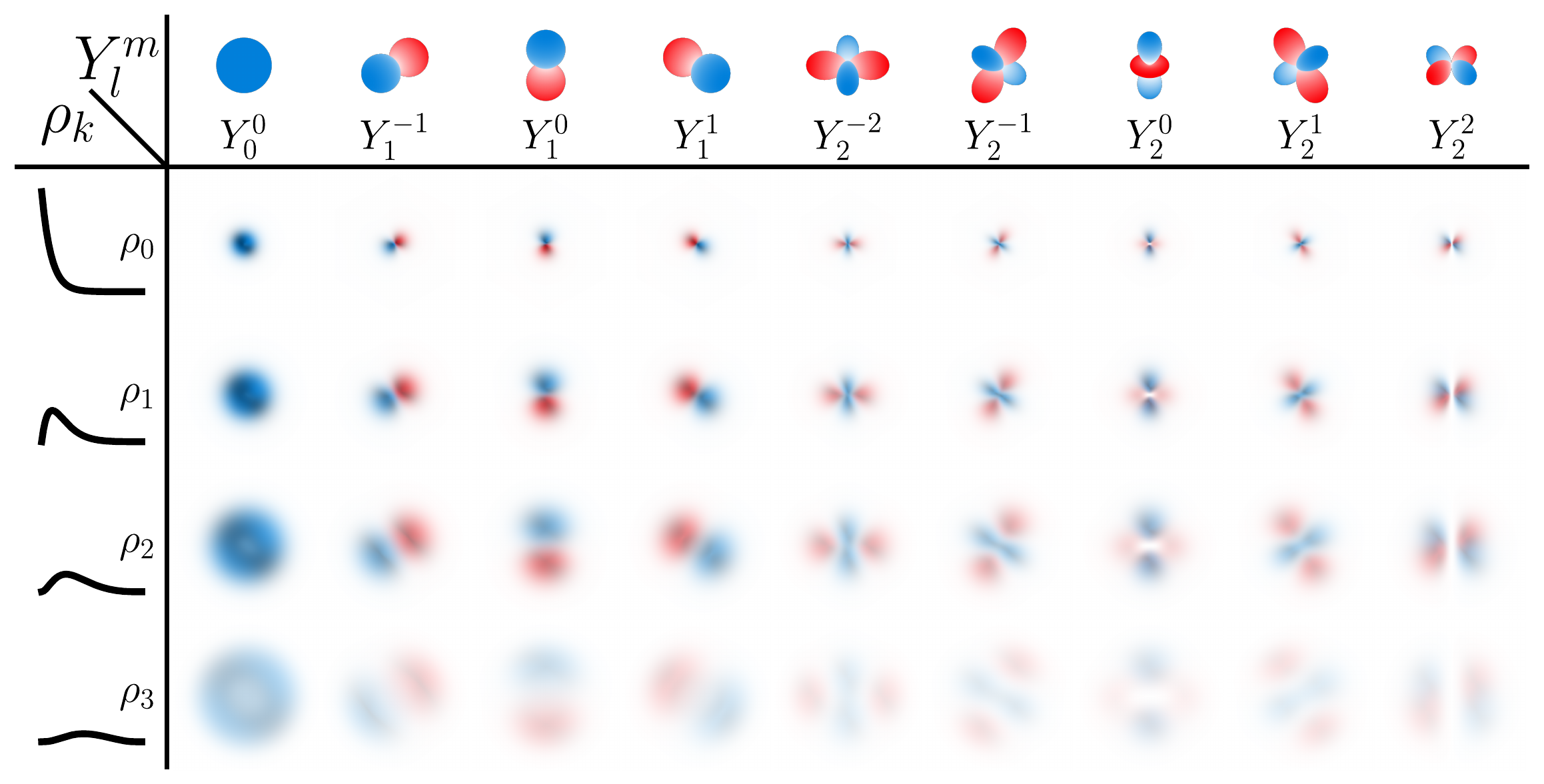}
	\caption{Visualizations of all basis functions $^{\protect\vphantom{0}}_{k}\mathrm{g}_{l}^{m}$ with $K=4$ (see Eq.~\ref{eq:basis_functions}) with different radial and angular components $\rho_k$ (Eq.~\ref{eq:bernstein_basis_function}) and $Y_l^m$ (Eq.~\ref{eq:spherical_harmonics}).}
	\label{fig:basis_functions}
\end{figure}
The basis functions (see Fig.~\ref{fig:basis_functions}) are given by
\begin{equation}
\begin{aligned}
\mathbf{g}_\mathrm{s}(\vv{r}) &= \begin{bmatrix}
^{\vphantom{0}}_{0}\mathrm{g}_{0}^{0}\\
\vdots \\
^{\vphantom{0}}_{K-1}\mathrm{g}_{0}^{0}
\end{bmatrix}\,,\\
\vv{\mathbf{g}}_\mathrm{p}(\vv{r}) &= \begin{bmatrix}
^{\vphantom{0}}_{0}\mathrm{g}_{1}^{-1} & ^{\vphantom{0}}_{0}\mathrm{g}_{1}^{0} & ^{\vphantom{0}}_{0}\mathrm{g}_{1}^{1} \\
\vdots & \vdots & \vdots \\
^{\vphantom{0}}_{K-1}\mathrm{g}_{1}^{-1} & ^{\vphantom{0}}_{K-1}\mathrm{g}_{1}^{0} & ^{\vphantom{0}}_{K-1}\mathrm{g}_{1}^{1}
\end{bmatrix} \,,
\\
\vv{\mathbf{g}}_\mathrm{d}(\vv{r}) &= \begin{bmatrix}
^{\vphantom{0}}_{0}\mathrm{g}_{2}^{-2} & ^{\vphantom{0}}_{0}\mathrm{g}_{2}^{-1} & ^{\vphantom{0}}_{0}\mathrm{g}_{2}^{0} & ^{\vphantom{0}}_{0}\mathrm{g}_{2}^{1} & ^{\vphantom{0}}_{0}g_{2}^{2} \\
\vdots & \vdots & \vdots & \vdots & \vdots \\
^{\vphantom{0}}_{K-1}\mathrm{g}_{2}^{-2} & ^{\vphantom{0}}_{K-1}\mathrm{g}_{2}^{-1} & ^{\vphantom{0}}_{K-1}\mathrm{g}_{2}^{0} & ^{\vphantom{0}}_{K-1}\mathrm{g}_{2}^{1} & ^{\vphantom{0}}_{K-1}\mathrm{g}_{2}^{2}
\end{bmatrix} \,,
\\
^{\vphantom{0}}_{k}\mathrm{g}_{l}^{m} &= \rho_k(\lVert\vv{r}\rVert)\cdot Y_{l}^{m}(\vv{r})\,,
\end{aligned}
\label{eq:basis_functions}
\end{equation}
where the radial component $\rho_k$ is
\begin{equation}
\rho_k(r) = b_{k,K-1}\left(\exp({-\gamma r})\right)\cdot f_{\rm cut}(r)
\label{eq:bernstein_basis_function}
\end{equation}
and
\begin{equation}
b_{k,K-1}(x) =\binom{K-1}{k}x^{k}(1-x)^{K-1-k}
\label{eq:bernstein_polynomials}
\end{equation}
are Bernstein polynomials ($k=0,\dots, K-1$). The hyper-parameter $K$ determines the total number of radial components (and the degree of the Bernstein polynomials). For $K\to\infty$, linear combinations of $b_{k,K-1}(x)$ can approximate any continuous function on the interval $[0,1]$ uniformly.\cite{bernstein1912demo} An exponential function $\exp({-\gamma r})$ maps distances $r$ from $[0,\infty)$ to the interval $(0,1]$, where $\gamma\in\mathbb{R}_{>0}$ is a radial decay parameter shared across all basis functions (for computational efficiency). A desirable side effect of this mapping is that the rate at which learned interaction functions can vary decreases with increasing~$r$, which introduces a chemically meaningful inductive bias (electronic wave functions also decay exponentially with increasing distance from a nucleus).\cite{unke2019physnet,hermann2020deep} The cutoff function
\begin{equation}
f_{\rm cut}(r) = \begin{cases}
\exp\left(-\dfrac{r^2}{(r_{\rm cut}-r)(r_{\rm cut}+r)}\right) & r < r_{\rm cut}\\
0 & r \geq r_{\rm cut}
\end{cases}
\label{eq:cutoff_function}
\end{equation}
ensures that basis functions smoothly decay to zero for $r\geq r_{\rm cut}$, so that no discontinuities are introduced when atoms enter or leave the cutoff radius. The angular component $Y_{l}^{m}(\vv{r})$ in Eq.~\ref{eq:basis_functions} is given by
\begin{equation}
\begin{aligned}
Y_{l}^{m}(\vv{r}) &= \begin{cases}
\sqrt{2}\cdot \Pi_l^{\lvert m\rvert}(z) \cdot A_{\lvert m \rvert}(x,y)
& m < 0 \\
\Pi_l^{0}(z)
& m = 0 \\
\sqrt{2}\cdot \Pi_l^{m}(z) \cdot B_m(x,y)
& m > 0 \\
\end{cases}\,,\\
A_{m}(x,y) &= \sum_{p=0}^{ m}\binom{m}{p}x^{p} y^{m-p} \sin\left(\frac{\pi}{2}( m-p)\right)\,,\\
B_{m}(x,y) &= \sum_{p=0}^{m}\binom{m}{p}x^{p} y^{m-p} \cos\left(\frac{\pi}{2}(m-p)\right)\,,\\
\Pi_{l}^{m}(z) &= \sqrt{\frac{(l-m)!}{(l+m)!}} \sum_{p=0}^{\lfloor(l-m)/2\rfloor} \ c_{plm} r^{2p-l}z^{l-2p-m}\,,\\
c_{plm} &= \frac{(-1)^p}{2^l}\binom{l}{p}\binom{2l-2p}{l}\frac{(l-2p)!}{(l-2p-m)!}\,,
\end{aligned}
\label{eq:spherical_harmonics}
\end{equation}
where $\vv{r} = [x\ y \ z]^{\mathsf{T}}$ and $r=\lVert\vv{r}\rVert$. Note that the $Y_{l}^{m}$ in Eq.~\ref{eq:spherical_harmonics} omit the normalization constant $\sqrt{(4\pi)/(2l+1)}$, but are otherwise identical to the standard (real) spherical harmonics.

Although locality is a valid assumption for many chemical systems,\cite{unke2020machine} electrons may also be delocalized across multiple distant atoms. Starting from the temporary features $\mathbf{\tilde{x}}$ (see Eq.~\ref{eq:interaction_module}), such nonlocal interactions are modeled via \emph{self-attention}\cite{vaswani2017attention} as
\begin{equation}
\begin{aligned}
\mathbf{q}_i &= \mathrm{resmlp}_q(\tilde{\mathbf{x}}_i)\,, &\mathbf{Q} = \left[\mathbf{q}_1 \ \cdots \ \mathbf{q}_N\right]^\mathsf{T}\,, \\
\mathbf{k}_i &= \mathrm{resmlp}_k(\tilde{\mathbf{x}}_i)\,, &\mathbf{K} = \left[\mathbf{k}_1 \ \cdots \ \mathbf{k}_N\right]^\mathsf{T}\,, \\
\mathbf{v}_i &= \mathrm{resmlp}_v(\tilde{\mathbf{x}}_i)\,, &\mathbf{V} = \left[\mathbf{v}_1 \ \cdots \ \mathbf{v}_N\right]^\mathsf{T}\,, \\
\mathbf{N} &= \mathrm{attention}(\mathbf{Q}, \mathbf{K}, \mathbf{V})\,, \quad  &\mathbf{N} = \left[\mathbf{n}_1 \ \cdots \ \mathbf{n}_N\right]^\mathsf{T}\,,
\end{aligned}
\label{eq:nonlocal_interaction}
\end{equation}
where the features $\mathbf{n}$ in Eq.~\ref{eq:interaction_module} are the (transposed) rows of the matrix $\mathbf{N}\in\mathbb{R}^{N\times F}$. The idea of attention is inspired by retrieval systems,\cite{kowalski2007information} where a \emph{query} is mapped against \emph{keys} to retrieve the best-matched corresponding \emph{values} from a database. Standard attention is computed as
\begin{equation}
\begin{aligned}
&\mathbf{A} = \exp\left({\mathbf{Q}\mathbf{K}^{\mathsf{T}}/\sqrt{F}}\right)\,,\quad \mathbf{D}=\mathrm{diag}(\mathbf{A}\mathbf{1}_N)\,,\\
&\mathrm{attention}(\mathbf{Q},\mathbf{K}, \mathbf{V}) = \mathbf{D}^{-1}\mathbf{A}\mathbf{V}\,,
\end{aligned}
\label{eq:attention}
\end{equation}
where $\mathbf{Q},\mathbf{K},\mathbf{V}\in \mathbb{R}^{N\times F}$ are queries, keys, and values, $\mathbf{1}_N$ is the all-ones vector of length $N$, and $\mathrm{diag}(\cdot)$ is a diagonal matrix with the input vector as the diagonal. Unfortunately, computing attention with Eq.~\ref{eq:attention} has a time and space complexity of $\mathcal{O}(N^2F)$ and $\mathcal{O}(N^2+NF)$,\cite{choromanski2020rethinking} respectively, because the attention matrix $\mathbf{A}\in\mathbb{R}^{N\times N}$ has to be stored explicitly. Since quadratic scaling with the number of atoms $N$ is problematic for large chemical systems, the FAVOR+ (\underline{F}ast \underline{A}ttention \underline{V}ia positive \underline{O}rthogonal \underline{R}andom features) approximation\cite{choromanski2020rethinking} is used instead:
\begin{equation}
\begin{aligned}
&\widehat{\mathbf{Q}} = \left[\phi(\mathbf{q}_i) \ \cdots \ \phi(\mathbf{q}_N)\right]^{\mathsf{T}}\,,\quad \widehat{\mathbf{K}} = \left[\phi(\mathbf{k}_i) \ \cdots \ \phi(\mathbf{k}_N)\right]^{\mathsf{T}}\,,\\
&\widehat{\mathbf{D}}=\mathrm{diag}\left(
\widehat{\mathbf{Q}}\left(\widehat{\mathbf{K}}^{\mathsf{T}}\mathbf{1}_N\right)\right)\,,\\
&\mathrm{attention}(\mathbf{Q},\mathbf{K}, \mathbf{V}) = \widehat{\mathbf{D}}^{-1}(\widehat{\mathbf{Q}}(\widehat{\mathbf{K}}^{\mathsf{T}}\mathbf{V}))\,.
\end{aligned}
\label{eq:performer_attention}
\end{equation}
Here $\phi:\mathbb{R}^{F}\mapsto\mathbb{R}_{>0}^{f}$ is a mapping designed to approximate the softmax kernel via $f$ random features, see Ref.~\citenum{choromanski2020rethinking} for details (here, $f=F$ for simplicity). The time and space complexities for computing attention with Eq.~\ref{eq:performer_attention} are $\mathcal{O}(NFf)$ and $\mathcal{O}(NF+Nf+Ff)$,\cite{choromanski2020rethinking} i.e.\ both scale linearly with the number of atoms $N$. To make the evaluation of \nn{} deterministic, the random features of the mapping $\phi$ are drawn only once at initialization and kept fixed afterwards (instead of redrawing them for each evaluation).\\
%Note that attention allows information to be communicated between all atoms in the system, regardless of their distance.\\

Once all interaction modules are evaluated, atomic energy contributions~$E_i$ are predicted from the atomic descriptors~$\mathbf{f}_i$ via linear regression
\begin{equation}
E_i = \mathbf{w}_E^{\mathsf{T}}\mathbf{f}_i + \tilde{E}_{Z_i}\,,
\label{eq:atomic_energy}
\end{equation}
and combined to obtain the total potential energy (see Eq.~\ref{eq:potential_energy}).
Here, $\mathbf{w}_E \in \mathbb{R}^F$ are the regression weights and $\tilde{E}_{Z}\in\mathbb{R}$ are element-dependent energy biases.

The nuclear repulsion term $E_{\rm rep}$ in Eq.~\ref{eq:potential_energy} is based on the Ziegler-Biersack-Littmark stopping potential\cite{ziegler1985stopping} and given by
\begin{equation}
\begin{split}
E_{\rm rep} &=  k_e  \sum_{i}\sum_{j>i\,\in\mathcal{N}(i)}  \frac{Z_iZ_j}{r_{ij}} f_{\rm cut}(r_{ij}) \cdot \\
&\qquad \left(\sum_{k=1}^{4} c_ke^{-a_k r_{ij} (Z_i^p+Z_j^p)/d}\right)\,.
\end{split}
\label{eq:repuslion_term}
\end{equation}
Here, $k_e$ is the Coulomb constant and $a_k$, $c_k$, $p$, and $d$ are parameters (see Eqs.~\ref{eq:local_interaction}~and~\ref{eq:cutoff_function} for the definitions of $\mathcal{N}(i)$ and $f_{\rm cut}$). Long-range electrostatic interactions are modeled as
\begin{equation}
E_{\rm ele} = k_e  \sum_{i}\sum_{j>i} q_iq_j\left(\frac{f_{\rm switch}(r_{ij})}{\sqrt{r_{ij}^2+1}} + \frac{1-f_{\rm switch}(r_{ij})}{r}\right)\,,
\label{eq:electrostatic_term}
\end{equation}
where $q_i$ are atomic partial charges predicted from the atomic features $\mathbf{f}_i$ according to
\begin{equation}
q_i = \mathbf{w}_q^{\mathsf{T}}\mathbf{f}_i + \tilde{q}_{Z_i} + \frac{1}{N}\left[Q-\sum_{j=1}^N\left(\mathbf{w}_q^{\mathsf{T}}\mathbf{f}_j + \tilde{q}_{Z_j}\right)\right]\,.
\label{eq:partial_charges}
\end{equation}
Here, $\mathbf{w}_q \in \mathbb{R}^F$ and $\tilde{q}_{Z}\in \mathbb{R}$ are regression weights and element-dependent biases, respectively. The second half of the equation ensures that $\sum_i q_i = Q$, i.e.\ the total charge is conserved. Standard Ewald summation\cite{ewald1921berechnung} can be used to evaluate $E_{\rm ele}$ when periodic boundary conditions are used.
Note that Eq.~\ref{eq:electrostatic_term} smoothly interpolates between the correct $r^{-1}$ behavior of Coulomb's law at large distances ($r > r_{\rm off}$) and a damped $(r_{ij}^2+1)^{-1/2}$ dependence at short distances ($r < r_{\rm on}$) via a smooth switching function $f_{\rm switch}$ given by
\begin{equation}
\begin{aligned}
\sigma(x) &= \begin{cases}
\exp\left(-\frac{1}{x}\right)& x > 0\\
0 & x \leq 0
\end{cases}\,,\\
f_{\rm switch}(r) &= \frac{\sigma\left(1-\frac{r-r_{\rm on}}{r_{\rm off}-r_{\rm on}}\right)}{\sigma\left(1-\frac{r-r_{\rm on}}{r_{\rm off}-r_{\rm on}}\right) +  \sigma\left(\frac{r-r_{\rm on}}{r_{\rm off}-r_{\rm on}}\right)}\,.
\end{aligned}
\label{eq:switch_function}
\end{equation}
For simplicity, $r_{\rm on} = \frac{1}{4}r_{\rm cut}$ and $r_{\rm off} = \frac{3}{4}r_{\rm cut}$, i.e.\ the switching interval is automatically adjusted depending on the chosen cutoff radius $r_{\rm cut}$ (see Eq.~\ref{eq:cutoff_function}). It is also possible to construct dipole moments~$\vv{\mu}$ from the partial charges according to
\begin{equation}
\vv{\mu} = \sum_{i=1}^{N} q_i \vv{r}_i\,,
\label{eq:dipole}
\end{equation}
which can be useful for calculating infrared spectra from MD simulations and for fitting $q_i$ to \textit{ab initio} reference data without imposing arbitrary charge decomposition schemes.\cite{gastegger2017machine}  Long-range dispersion interactions are modeled via the term $E_{\rm vdw}$. Analytical van der Waals corrections are an active area of research and many different methods, for example the Tkatchenko-Scheffler correction,\cite{tkatchenko2009accurate} or many body dispersion,\cite{tkatchenko2012accurate} have been proposed.\cite{hermann2017first} In this work, the two-body term of the D4 dispersion correction\cite{caldeweyher2019generally} is used for its simplicity and computational efficiency:
\begin{equation}
E_{\rm vdw} = - \sum_{i}\sum_{j>i} \sum_{n=6,8}s_n \frac{C^{ij}_{(n)}}{r_{ij}^n}f_{\rm damp}^{(n)}(r_{ij})\,.
\label{eq:dispersion_term}
\end{equation}
Here $s_n$ are scaling parameters, $f_{\rm damp}^{(n)}$ is a damping function, and $C^{ij}_{(n)}$ are pairwise dispersion coefficients. They are obtained by interpolating between tabulated reference values based on a (geometry-dependent) fractional coordination number and an atomic partial charge $q_i$. In the standard D4 scheme, the partial charges are obtained via a charge equilibration scheme,\cite{caldeweyher2019generally} in this work, however, the $q_i$ from Eq.~\ref{eq:partial_charges} are used instead. Note that the D4 method was developed mainly to correct for the lack of dispersion in density functionals, so typically, some of its parameters are adapted to the functional the correction is applied to (optimal values for each functional are determined by fitting to high-quality electronic reference data).\cite{caldeweyher2019generally} In this work, all D4 parameters that vary between different functionals are treated as learnable parameters when \nn{} is trained, i.e.\ they are automatically adapted to the reference data. Since Eq.~\ref{eq:partial_charges} (instead of charge equilibration) is used to determine the partial charges, an additional learnable parameter $s_q$ is introduced to scale the tabulated reference charges used to determine dispersion coefficients $C^{ij}_{(n)}$. For further details on the implementation of the D4 method, the reader is referred to Ref.~\citenum{caldeweyher2019generally}.\\

\subsection*{Training and hyperparameters}
All \nn{} models in this work use $T=6$ interaction modules, $F=128$ features, and a cutoff radius $r_\mathrm{cut} = 10$~$a_0$ ($\approx$5.29177~\AA), unless otherwise specified. Weights are initialized as random (semi\nobreakdash-)orthogonal matrices with entries scaled according to the Glorot initialization scheme.\cite{glorot2010understanding}
An exception are the weights of the second linear layer in residual blocks ($\mathrm{linear}_2$ in Eq.~\ref{eq:residual}) and the matrix $\mathbf{M}$ used in nuclear embeddings (Eq.~\ref{eq:nuclear_embedding}), which are initialized with zeros. All bias terms and the $\mathbf{\tilde{k}}$ and $\mathbf{\tilde{v}}$ parameters in the electronic embedding (Eq.~\ref{eq:electronic_embedding}) are also initialized with zeros. The parameters for the activation function (Eq.~\ref{eq:activation}) start as $\alpha = 1.0$ and $\beta=1.702$, following the recommendations given in Ref.~\citenum{hendrycks2016gaussian}. The radial decay parameter $\gamma$ used in Eq.~\ref{eq:bernstein_basis_function} is initialized to $\frac{1}{2}$~$a_0^{-1}$ and constrained to positive values. The parameters of the empirical nuclear repulsion term (Eq.~\ref{eq:repuslion_term}) start from the literature values of the ZBL potential\cite{ziegler1985stopping} and are constrained to positive values (coefficients $c_k$ are further constrained such that $\sum c_k = 1$ to guarantee the correct asymptotic behavior for short distances). Parameters of the dispersion correction (Eq.~\ref{eq:dispersion_term}) start from the values recommended for Hartree-Fock calculations\cite{caldeweyher2019generally} and the charge scaling parameter $s_q$ is initialized to $1$ (and constrained to remain positive).

The parameters are trained by minimizing a loss function
with mini-batch gradient descent using the AMSGrad optimizer\cite{reddi2019convergence}  with the recommended default momentum hyperparameters and an initial learning rate of 10$^{-3}$. During training, an exponential moving average of all model parameters is kept using a smoothing factor of 0.999. Every 1k training steps, a model using the averaged parameters is evaluated on the validation set and the learning rate is decayed by a factor of 0.5 whenever the validation loss does not decrease for 25 consecutive evaluations. Training is stopped when the learning rate drops below 10$^{-5}$ and the model that performed best on the validation set is selected. The loss function is given by
\begin{equation}
\mathcal{L} = \alpha_E\mathcal{L}_E + \alpha_F\mathcal{L}_F+ \alpha_\mu\mathcal{L}_\mu\,,
\label{eq:loss}
\end{equation}
where $\mathcal{L}_E$, $\mathcal{L}_F$, and $\mathcal{L}_\mu$ are separate loss terms for energies, forces and dipole moments and $\alpha_E$, $\alpha_F$, and $\alpha_\mu$ corresponding weighting hyperparameters that determine the relative influence of each term to the total loss. The energy loss is given by
\begin{equation}
\mathcal{L}_E =\sqrt{\frac{1}{B}\sum_{b=1}^{B} \left(E_{{\rm pot},b}-E_{{\rm pot},b}^{\rm ref}\right)^2}\,,
\label{eq:energy_loss}
\end{equation}
where $B$ is the number of structures in the mini-batch, $E_{{\rm pot},b}$ the predicted potential energy (Eq.~\ref{eq:potential_energy}) for structure~$b$ and $E_{{\rm pot},b}^{\rm ref}$ the corresponding reference energy. The batch size~$B$ is chosen depending on the available training data: When training sets contain 1k structures or less, $B=1$, for 10k structures or less, $B=10$, and for more than 10k structures, $B=100$. The force loss is given by
\begin{equation}
\mathcal{L}_F =\sqrt{\frac{1}{B}\sum_{b=1}^{B}\left(\frac{1}{N_b}\sum_{i=1}^{N_b}\left\lVert{-\frac{\partial E_{{\rm pot,b}}}{\partial \vv{r}_{i,b}} - \vv{F}_{i,b}^{\rm ref}}\right\rVert^2\right)}\,,
\label{eq:forces_loss}
\end{equation}
where $N_b$ is the number of atoms in structure~$b$ and $\vv{F}_{i,b}^{\rm ref}$ the reference force acting on atom~$i$ in structure~$b$. The dipole loss
\begin{equation}
\mathcal{L}_\mu =\sqrt{\frac{1}{B}\sum_{b=1}^{B} \left\lVert\left(\sum_{i=1}^{N_b}q_{i,b}\vv{r}_{i,b}\right)-\vv{\mu}_b^{\rm ref}\right\rVert^2}
\label{eq:dipole_loss}
\end{equation}
allows to learn partial charges (Eq.~\ref{eq:partial_charges}) from reference dipole moments $\vv{\mu}_b^{\rm ref}$, which are, in contrast to arbitrary charge decompositions, true quantum mechanical observables.\cite{gastegger2017machine} Note that for charged molecules, the dipole moment is dependent on the origin of the coordinate system, so consistent conventions must be used.
For some data sets or applications, however, reference partial charges $q_{i,b}^{\rm ref}$ obtained from a charge decomposition scheme, e.g.\ Hirshfeld charges,\cite{hirshfeld1977bonded} might be preferred (or the only data available). In this case, the term $\alpha_\mu\mathcal{L}_\mu$ in Eq.~\ref{eq:loss} is replaced by $\alpha_q\mathcal{L}_q$ with
\begin{equation}
\mathcal{L}_q =\sqrt{\frac{1}{B}\sum_{b=1}^{B} \left(\frac{1}{N_b}\sum_{i=1}^{N_b}\left(q_{i,b}-q_{i,b}^{\rm ref}\right)^2\right)}\,.
\label{eq:charge_loss}
\end{equation}

For simplicity, the relative loss weights are set to $\alpha_E=\alpha_F=\alpha_{\mu/q}=1$ in this work, with the exception of the MD17 and QM7-X data sets, for which $\alpha_F=100$ is used following previous work.\cite{unke2019physnet} Both energy and force prediction errors are significantly reduced when the force weight is increased (see Table~\ref{Stab:qm7x_results_low_force_weight}). Note that the relative weight of loss terms also depends on the chosen unit system (atomic units are used here). For data sets that lack the reference data necessary for computing any of the given loss terms (Eqs.~\ref{eq:energy_loss}--\ref{eq:charge_loss}), the corresponding weight is set to zero. In addition, whenever no reference data (neither dipole moments nor reference partial charges) are available to fit partial charges, both $E_{\rm ele}$ and $E_{\rm vdw}$ are omitted when predicting the potential energy $E_{\rm pot}$ (see Eq.~\ref{eq:potential_energy}).

For the ``unknown molecules/unknown conformations'' task reported in Table~\ref{tab:qm7x_results}, the 25 entries with the following ID numbers (\texttt{idmol} field in the QM7-X file format) were used as a test set: \texttt{1771}, \texttt{1805}, \texttt{1824}, \texttt{2020}, \texttt{2085}, \texttt{2117}, \texttt{3019}, \texttt{3108}, \texttt{3190}, \texttt{3217}, \texttt{3257}, \texttt{3329}, \texttt{3531}, \texttt{4010}, \texttt{4181}, \texttt{4319}, \texttt{4713}, \texttt{5174}, \texttt{5370}, \texttt{5580}, \texttt{5891}, \texttt{6315}, \texttt{6583}, \texttt{6809}, \texttt{7020}. Note, that in addition to energies and forces, \nn{} uses dipole moments (property $D$ in the QM7-X dataset) to fit atomic partial charges.\\

\subsection*{Computing and visualizing local chemical potentials and nonlocal contributions}
To compute the local chemical potentials shown in Figs.~\ref{fig:insights}a~and~\ref{Sfig:different_spin_gummybears}, a similar approach as that described in Ref.~\citenum{schutt2017quantum} is followed. To compute the local chemical potential $\Omega_A^M(\vv{r})$ of a molecule $M$ for an atom of type $A$ (here, hydrogen is used), the idea is to introduce a probe atom of type $A$ at position $\vv{r}$ and let it interact with all atoms of $M$, but not \textit{vice versa}. In other words, the prediction for $M$ is unperturbed, but the probe atom ``feels'' the presence of $M$. Then, the predicted energy contribution of the probe atom is interpreted as its local chemical potential $\Omega_A^M(\vv{r})$. This is achieved as follows: First, the electronic embeddings (Eq.~\ref{eq:electronic_embedding}) for all $N$ atoms in $M$ are computed as if the probe atom was not present. Then, the embeddings for the probe atom are computed as if it was part of a larger molecule with $N+1$ atoms. Similarly, the contributions of local interactions (Eq.~\ref{eq:local_interaction}) and nonlocal interactions (Eq.~\ref{eq:nonlocal_interaction}) to the features of the probe atom are computed by pretending it is part of a molecule with $N+1$ atoms, whereas all contributions to the features of the $N$ atoms in molecule $M$ are computed without the presence of the probe atom. For visualization, all chemical potentials are projected onto the $\sum_{i=1}^{N}\lVert \vv{r}-\vv{r}_i \rVert^{-2} = 1~a_0^{-2}$ isosurface, where the sum runs over the positions $\vv{r}_i$ of all atoms $i$ in $M$.

To obtain the individual contributions for s-, p-, and d-orbital-like interactions shown in Fig.~\ref{fig:insights}, different terms for the computation of $\mathbf{l}$ in Eq.~\ref{eq:local_interaction} are set to zero. For the s-orbital-like contribution, both $\vv{\mathbf{p}}$ and $\vv{\mathbf{d}}$ are set to zero. For the p-orbital-like contribution, only $\vv{\mathbf{d}}$ is set to zero, and the s-orbital-like contribution is subtracted from the result. Similarly, for the d-orbital-like contribution, the model is evaluated normally and the result from setting only $\vv{\mathbf{d}}$ to zero is subtracted.

The nonlocal contributions to the potential energy surface shown in Fig.~\ref{fig:insights}b are obtained by first evaluating the model normally and then subtracting the predictions obtained when setting $\mathbf{n}$ in Eq.~\ref{eq:interaction_module} to zero.

\subsection*{SchNet and PaiNN training}
The SchNet and PaiNN models for the QM7-X experiments use $F=128$ features, as well as $T=6$ and $T=3$ interactions, respectively.
Both employ 20 Gaussian radial basis function up to a cutoff of 5~\AA. They were trained with the Adam optimizer~\cite{kingma2014adam} at a learning rate of $10^{-4}$ and a batch size of 10.\\

\subsection*{Data generation}
For demonstrating the ability of \nn{} to model different electronic states and nonlocal interactions, energies, forces, and dipoles for three new data sets were computed at the semi-empirical GFN2-xTB level of theory.\cite{bannwarth2019gfn2} Both the Ag$_3^+$/Ag$_3^-$ (see Fig.~\ref{fig:electronic_states}a) and the singlet/triplet CH$_2$ (see Fig.~\ref{fig:electronic_states}b) data sets were computed by sampling 550 structures around the minima of both electronic states with normal mode sampling\cite{smith2017ani} at 1000~K. Then, each sampled structure was re-computed in the other electronic state (e.g.\ all structures sampled for Ag$_3^+$ were re-computed with a negative charge), leading to a total of 2200 structures for each data set (models were trained on a subset of 1000 randomly sampled structures).

The data set for Fig.~\ref{fig:nonlocal} was computed by performing bond scans for all nine shown diatomic molecules using 1k points spaced evenly between $1.5$--$20$~$a_0$, leading to a total of 9k structures. Models were trained on all data with an increased cutoff  $r_{\rm cut} = 18$~$a_0$ to demonstrate that a model without nonlocal interactions is unable to fit the data, even when it is allowed to overfit and use a large cutoff.\\

\section*{Data and code availability}
The data sets generated for this work and a reference implementation of \nn{} using PyTorch\cite{paszke2019pytorch} will be made available when the manuscript is accepted. All other data sets used in this work are publicly available from Ref.~\citenum{ch4} (completeness test in Section~\ref{Ssec:atomic_descriptors}), \url{http://www.sgdml.org} (MD17),  Ref.~\citenum{qm7x} (QM7-X), Ref.~\citenum{behler} (data sets used in Table~\ref{tab:behler_results}), and Ref.~\citenum{qmspin} (QMspin).

\begin{acknowledgments}
OTU acknowledges funding from the Swiss National Science
Foundation (Grant No. P2BSP2\_188147). We thank the authors of Ref.~\citenum{pozdnyakov2020incompleteness} for sharing raw data for reproducing the learning curves shown in Fig.~\ref{Sfig:ceriotti_ch4_learning_curves} and the geometries displayed in Fig.~\ref{Sfig:descriptors}e.
KRM was supported in part by the Institute of Information \& Communications Technology Planning \& Evaluation (IITP) grant funded by the Korea Government (No. 2019-0-00079,  Artificial Intelligence Graduate School Program, Korea University), and was partly supported by the German Ministry for Education and Research (BMBF) under Grants 01IS14013A-E, 01GQ1115, 01GQ0850, 01IS18025A and 01IS18037A; the German Research Foundation (DFG) under Grant Math+, EXC 2046/1, Project ID 390685689. Correspondence should be addressed to OTU and KRM.
We thank Alexandre Tkatchenko and Hartmut Maennel for very helpful discussions and feedback on the manuscript.
\end{acknowledgments}

\bibliography{references}

\end{document}

%opening
\title{Supporting Information for \nn{}: Learning Force Fields with \\Electronic Degrees of Freedom and Nonlocal Effects}
\author{Oliver T. Unke}
\email{oliver.unke@googlemail.com}
\affiliation{Machine Learning Group, Technische Universit\"at Berlin, 10587 Berlin, Germany}
\affiliation{DFG Cluster of Excellence ``Unifying Systems in Catalysis'' (UniSysCat), Technische Universit\"at Berlin, 10623 Berlin, Germany}

\author{Stefan Chmiela}
\affiliation{Machine Learning Group, Technische Universit\"at Berlin, 10587 Berlin, Germany}

\author{Michael Gastegger}
\affiliation{Machine Learning Group, Technische Universit\"at Berlin, 10587 Berlin, Germany}
\affiliation{DFG Cluster of Excellence ``Unifying Systems in Catalysis'' (UniSysCat), Technische Universit\"at Berlin, 10623 Berlin, Germany}

\author{Kristof T. Sch\"utt}
\affiliation{Machine Learning Group, Technische Universit\"at Berlin, 10587 Berlin, Germany}

\author{Huziel E.\ Sauceda}
\affiliation{Machine Learning Group, Technische Universit\"at Berlin, 10587 Berlin, Germany}
\affiliation{
	BASLEARN, BASF-TU joint Lab, Technische Universit\"at Berlin, 10587 Berlin, Germany
}

\author{Klaus-Robert M\"uller}
\email{klaus-robert.mueller@tu-berlin.de}
\affiliation{Machine Learning Group, Technische Universit\"at Berlin, 10587 Berlin, Germany}
\affiliation{Department of Artificial Intelligence, Korea University, Anam-dong, Seongbuk-gu, Seoul 02841, Korea}
\affiliation{Max Planck Institute for Informatics, Stuhlsatzenhausweg, 66123 Saarbr\"ucken, Germany}
\affiliation{BIFOLD -- Berlin Institute for the Foundations of Learning and Data, Berlin, Germany}
\affiliation{Google Research, Brain team, Berlin, Germany}

\maketitle

\section{Completeness of atomic descriptors in \nn{}}
\label{sec:atomic_descriptors}
\begin{figure*}
	\includegraphics[width=\textwidth]{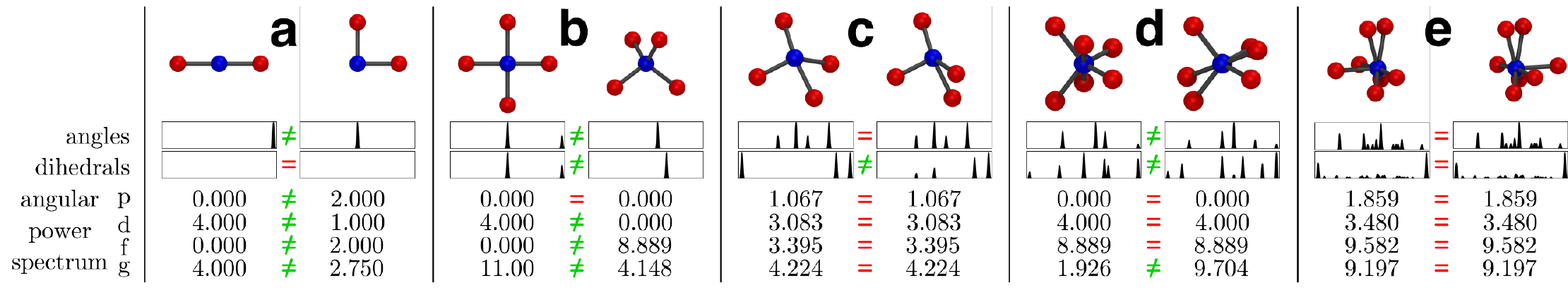}
	\caption{Pairs of distinct atomic environments where all neighboring atoms (red) have the same distance from the central atom (blue). The distributions of angles (Eq.~\ref{eq:angles}) and dihedrals (Eq.~\ref{eq:dihedrals}) are visualized and values of the angular power spectrum invariants (Eq.~\ref{eq:power_spectrum}) for different angular momenta $l=1,\dots4$ (p, d, f, g) are given for each structure (the s invariant simply counts the number of neighbors and is therefore omitted). Since all distances to neighboring atoms are identical, descriptors need to be able to at least resolve angular information to distinguish the structures (\textbf{a}). However, for some structures, the power spectrum invariants may be degenerate for small values of $l$ (\textbf{b} and \textbf{d}). Some structures even have identical angular distributions, in which case the power spectrum invariants are equal for all $l=0,\dots,\infty$ and information about dihedrals is necessary to distinguish the environments (\textbf{c}). Note that some environments cannot even be distinguished when information about dihedrals is included (\textbf{d}).\cite{pozdnyakov2020incompleteness}}
	\label{fig:descriptors}
\end{figure*}
Many ML algorithms for constructing potential energy surfaces make use of some sort of descriptor to represent atoms in their chemical environment. As long as this description is \emph{complete}, any  atom-centered property (including atomic decompositions of extensive properties such as energy) can be predicted from the descriptors.\cite{glielmo2018efficient} In this context, \emph{completeness} means that structures which are not convertible into each other (by translations, rotations, or permutations of equivalent atoms) map to different descriptors.

A simple descriptor for the environment of an atom~$i$ at position~$\vv{r}_{i}$ consists of the set of distances $r_{ij} = \lVert \vv{r}_{ij}\rVert$ (with $\vv{r}_{ij} = \vv{r}_{j}- \vv{r}_{i}$) to neighboring atoms $j$, and the set of angles
\begin{equation}
\alpha_{ijk} = \mathrm{arccos}\left(\frac{\langle \vv{r}_{ij}, \vv{r}_{ik} \rangle}{\lVert \vv{r}_{ij}\rVert \lVert \vv{r}_{ik}\rVert}\right)
\label{eq:angles}
\end{equation}
between all possible combinations of neighboring atoms $j$ and $k$. For environments consisting of multiple different species, separate sets of distances (and angles) are necessary for each element (or combination of elements). However, for simplicity, it is assumed here that all atoms are identical. A disadvantage of using angles in the descriptor is that their computation scales $\mathcal{O}(n^2)$ with the number of neighbors $n$, because all combinations must be considered. An alternative to encode angular information, which scales $\mathcal{O}(n)$, is to replace the set of angles with invariants of the form
\begin{equation}
a_{i,l} = \sum_{m=-l}^{l} \left(\sum_{j=1}^{n} Y_l^m(\vv{r}_{ij})\right)^2
\label{eq:power_spectrum}
\end{equation}
derived from the angular power spectrum, where $Y_l^m$ are the spherical harmonics (see Eq.~\ref{Meq:spherical_harmonics}). In the following, $a_{i,l}$ for $l=0,1,2,3,4$ are called s, p, d, f, and g invariants because of their relation to the symmetries of atomic orbitals. The disadvantage here is that when using a finite number ($l=0,\dots,L$) of power spectrum invariants as angular descriptor, some environments with different sets of angles may lead to the same descriptor. For example, square planar and terahedral environments have the same s and p invariants ($L=1$), so it is necessary to include at least d invariants ($L=2$) in the descriptor to differentiate them (see Fig.~\ref{fig:descriptors}B).\\

There is a widespread belief in the literature that sets of distances and angles are sufficient for a \emph{complete} description of atomic environments.\cite{von2015fourier,kocer2020continuous} However, it was recently demonstrated that this is not the case, and even including the set of dihedrals
\begin{equation}
\delta_{ijkl} = \mathrm{arccos}\left(\frac{\langle \vv{r}_{ij}\times\vv{r}_{ik}, \vv{r}_{ik}\times\vv{r}_{il} \rangle}{\lVert\vv{r}_{ij}\times\vv{r}_{ik}\rVert \lVert \vv{r}_{ik}\times\vv{r}_{il}\rVert}\right)
\label{eq:dihedrals}
\end{equation}
between triplets of neighboring atoms does not lead to a \textit{complete} description in general.\cite{pozdnyakov2020incompleteness} In this context, it is interesting to investigate the \emph{completeness} of the atomic descriptors~$\mathbf{f}$~(see Eq.~\ref{Meq:atomic_descriptor})  learned by \nn{} and compare its ability to distinguish different structures to other popular approaches. For this purpose, five pairs of distinct atomic environments (shown in Fig.~\ref{fig:descriptors}), with geometries that are particularly difficult to separate, are considered. Then, different models are trained to predict scalar labels of $1$ (for one of the environments) and $-1$ (for the other environment) from the descriptors of the central atoms (blue) in each pair. It can be observed that models either learn to predict the labels with virtually zero error (up to numerical precision), i.e.\ the environments can be distinguished, or a value of $0$ is predicted for both central atoms, i.e.\ their environments are mapped to the same descriptor and a compromise between the contradictory labels has to be found. The results are summarized in Table~\ref{tab:descriptors}. For evaluating PhysNet\cite{unke2019physnet} and DimeNet,\cite{klicpera2020directional} the reference implementations available from \url{https://github.com/MMunibas/PhysNet} and  \url{https://github.com/klicperajo/dimenet} are used. PaiNN\cite{schutt2021equivariant} and NequIP\cite{batzner2021se} are evaluated using in-house implementations. For BPNN and SchNet, the implementations available in SchNetPack\cite{schutt2018schnetpack} are used. The FCHL18/19\cite{faber2018alchemical,christensen2020fchl} models are evaluated using the QML package.\cite{qmlpackage} All models were trained on multiple randomly rotated versions of the environments shown in Fig.~\ref{fig:descriptors}. This was done to prevent models picking up on differences due to floating point imprecision, which otherwise may make environments distinguishable even when their descriptors are degenerate (up to numerical noise).
\begin{table}
	\begin{threeparttable}
		\begin{tabular} {l l c c c c c }
			\toprule
			\bf model & \bf scaling & \bf A & \bf B & \bf C & \bf D & \bf E \\
			\midrule
			\multicolumn{7}{c}{\bf hand-crafted descriptors}\\
			BPNN\cite{behler2007generalized} & $\mathcal{O}(n^2)$ & \cmark & \cmark & \xmark & \cmark & \xmark \\
			FCHL19\cite{christensen2020fchl} & $\mathcal{O}(n^2)$ & \cmark & \cmark & \xmark & \cmark & \xmark \\
			FCHL18\cite{faber2018alchemical} & $\mathcal{O}(n^3)$\tnote{*} & \cmark & \cmark & \cmark\tnote{**} & \cmark & \xmark \\
			\midrule
			\multicolumn{7}{c}{\bf learned descriptors}\\
			SchNet\cite{schutt2018schnet} & $\mathcal{O}(n)$ &(\cmark) & (\cmark) & (\cmark) & (\cmark) & (\cmark) \\
			PhysNet\cite{unke2019physnet} & $\mathcal{O}(n)$ & (\cmark) & (\cmark) & (\cmark) & (\cmark) & (\cmark) \\
			DimeNet\cite{klicpera2020directional} & $\mathcal{O}(n^2)$ & \cmark & \cmark & (\cmark) & \cmark & (\cmark) \\
			NequIP\cite{batzner2021se} & $\mathcal{O}(n)$ &\cmark & (\cmark) & (\cmark) & (\cmark) & (\cmark) \\
			PaiNN\cite{schutt2021equivariant} & $\mathcal{O}(n)$ & \cmark & (\cmark) & (\cmark) & (\cmark) & (\cmark) \\
			\nn{} & $\mathcal{O}(n)$ & \cmark & \cmark & (\cmark) & (\cmark) & (\cmark) \\
			\bottomrule
		\end{tabular}
		\begin{tablenotes}\footnotesize
			\item[*] when dihedrals are included
			\item[**] only distinguishable with dihedrals
		\end{tablenotes}
	\end{threeparttable}
	\caption{Ability of models to differentiate the atomic environments shown in Fig.~\ref{fig:descriptors} (\cmark:~distinguishable, \xmark:~indistinguishable) and the scaling of their computational cost  with respect to the number of neighbors $n$. Message-passing neural networks with learned descriptors can distinguish all environments when there are $T\geq2$ message-passing steps. However, when only a single step is used ($T=1$), the environments marked with (\cmark) become indistinguishable.}
	\label{tab:descriptors}
\end{table}

Most models based on hand-crafted descriptors can only distinguish environments when their sets of distances and angles (and in some cases dihedrals) differs. Message-passing neural networks (MPNNs) on the other hand can learn to distinguish all environments shown in Fig.~\ref{fig:descriptors}, provided that at least $T\geq2$ message-passing steps are used. The amount of information that can be resolved with a single message-passing step ($T=1$) is often related to the power spectrum invariants (see Eq.~\ref{eq:power_spectrum}) and different MPNNs mainly differ in the maximum order $L$ which they can resolve in a single update (with the exception of DimeNet,\cite{klicpera2020directional} which uses angles directly but scales $\mathcal{O}(n^2)$ with the number of neighbors $n$). \nn{} uses an update with a maximum order of $L=2$, which is sufficient to differentiate most common chemical environments (as long as they are distinguishable by distances and angles). It would be possible to introduce higher order interactions with the symmetry of f-, or even g-orbitals into the update step (see Eq.~\ref{Meq:local_interaction}), so that additional environments (e.g.\ Fig.~\ref{fig:descriptors}d) become distinguishable with a single update, but this increases the computational cost and is found to give little benefit (in terms of additional accuracy for predictions) in practice.\\

%It should be noted that even if a descriptor cannot distinguish all atomic environments, this does not necessarily mean that a model relying on this descriptor must fail when predicting extensive properties such as the potential energy. For example, even though the environments of the central atoms (blue) shown in Fig.~\ref{fig:descriptors}C cannot be distinguished with a descriptor based on just distances and angles, the descriptors of neighboring atoms (red) will differ between both structures. Consequently, when the potential energy is modeled as a sum over atomic contributions, it is possible to compensate for the degenerate atoms with the energy contributions predicted for their neighbors.

\begin{table*}
	\begin{tabular}{r c c c c}
		\toprule
		& \multicolumn{2}{c}{\textbf{energy} [kcal~mol$^{-1}$]} & \multicolumn{2}{c}{\textbf{forces} [kcal~mol$^{-1}$~\AA$^{-1}$]}\\ 
		$n_{\rm train}$ & MAE & RMSE & MAE & RMSE\\
		\midrule
		10 & \quad 11.698 (2.440) & \quad 18.650 (4.889) & \quad 14.426 (3.280) & \quad 40.302 (16.473) \\
		100 & \quad 4.011 (1.688) & \quad 6.183 (1.969) & \quad 5.782 (1.436) & \quad 14.345 (4.730) \\
		1\,000 & \quad 0.607 (0.030) & \quad 1.646 (0.304) & \quad 1.360	(0.039) & \quad 5.326 (2.614) \\
		10\,000 &  \quad 0.078 (0.002) &  \quad 0.282 (0.009) &  \quad 0.249 (0.007) &  \quad 0.998	(0.032)\\
		100\,000 &  \quad 0.020	(0.001) &  \quad 0.071	(0.003)	&  \quad 0.071 (0.001)	& \quad 0.326 (0.012)\\
		1\,000\,000 &  \quad 0.020	(0.002)	&  \quad  0.036	(0.003) &  \quad  0.063	(0.006)	&  \quad 0.165	(0.015)\\
		\bottomrule
	\end{tabular}
	\caption{Mean absolute errors (MAEs) and root mean square errors (RMSEs) of energies and forces for the random CH$_4$ dataset\cite{ch4} suggested in Ref.~\citenum{pozdnyakov2020incompleteness}. Results are averaged over 16 ($n_{\rm train} = 10$), 8 ($n_{\rm train} = 100$), or 4 ($n_{\rm train} \geq 1000$) random splits and the standard deviation between runs is given in brackets.}
	\label{tab:ceriotti_ch4_results}
\end{table*}

\begin{figure}
	\includegraphics[width=\columnwidth]{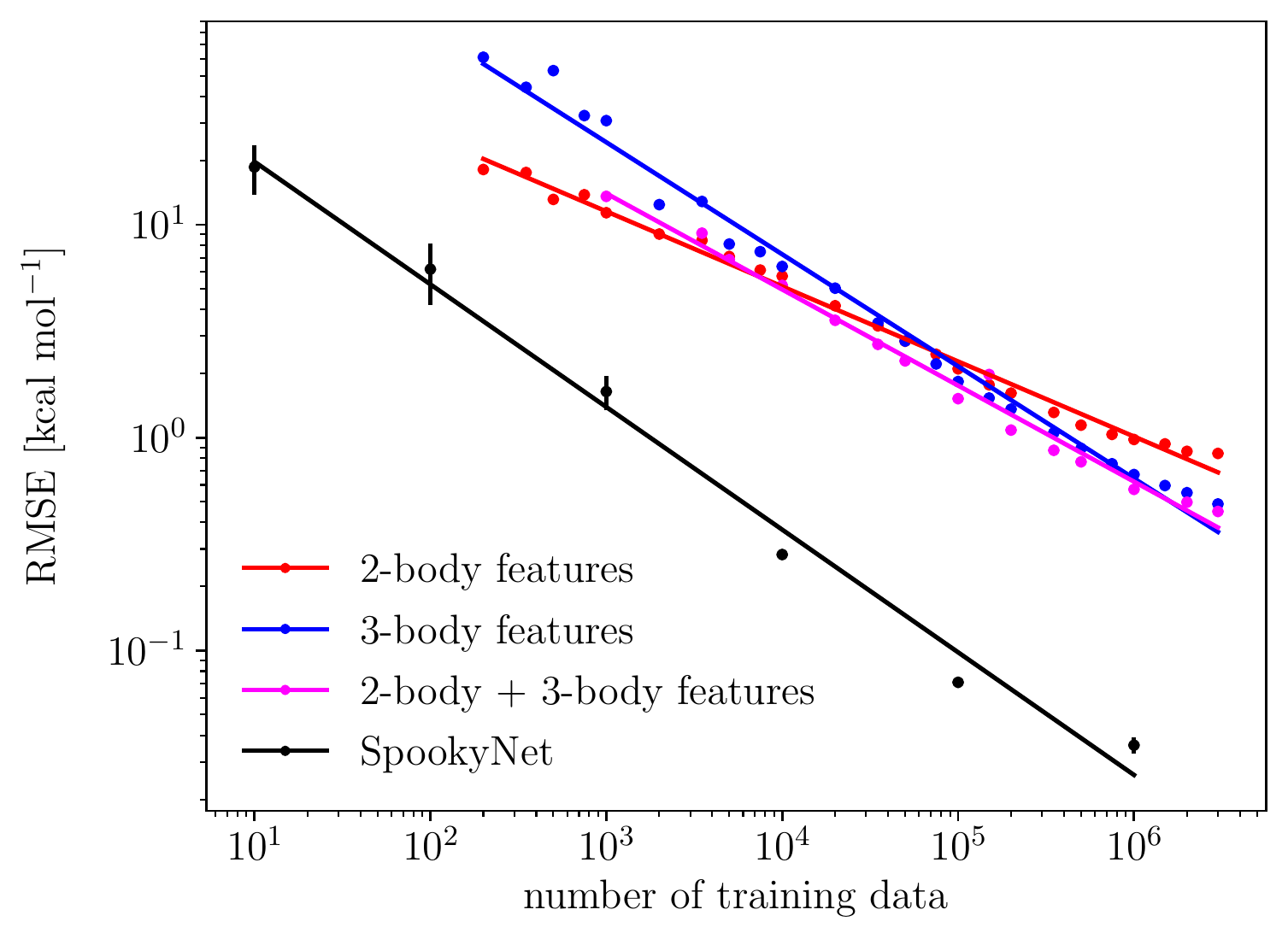}
	\caption{Energy learning curves for the random CH$_4$ dataset\cite{ch4} suggested in Ref.~\citenum{pozdnyakov2020incompleteness}. \nn{} (black) is compared to feedforward neural networks trained on 2-body and/or 3-body features (red, blue, magenta), see Ref.~\citenum{pozdnyakov2020incompleteness} for details.}
	\label{fig:ceriotti_ch4_learning_curves}
\end{figure}

\citeauthor{pozdnyakov2020incompleteness} propose a different \emph{completeness} test based on a data set of $\sim$7.7M CH$_4$ structures\cite{ch4} that were generated by randomly placing H atoms in a 3~\AA{} sphere around the C atom (see Ref.~\citenum{pozdnyakov2020incompleteness} for details). Due to the strongly distorted geometries, potential energies in this data set vary by $\sim$1400 kcal~mol$^{-1}$ and forces by $\sim$10700 kcal~mol$^{-1}$~\AA$^{-1}$. Further, this way of sampling will lead to many structures with (nearly) degenerate sets of angles (see Fig.~\ref{fig:descriptors}c) and is thus particularly challenging to learn. For this task, it is to be expected that models relying on \emph{incomplete} descriptors improve at a slower rate (and eventually cease to improve at all) when increasing the number of training data.\cite{pozdnyakov2020incompleteness} The performance of \nn{} on this data set for different training set sizes is summarized in Table~\ref{tab:ceriotti_ch4_results}. With only 10 training points ($\sim$0.00013\% of the data), \nn{} reaches prediction errors that correspond to a relative absolute error of just $\sim$1\% (with respect to the energy range covered in the data set). Chemical accuracy (absolute errors $<$1~kcal~mol$^{-1}$) is reached with as few as 1000 training points. The learning curve (see Fig.~\ref{fig:ceriotti_ch4_learning_curves}) shows that the performance of \nn{} increases steadily when more data is used for training while being about two orders of magnitude more data-efficient than other methods. 

\begin{figure*}
	\includegraphics[width=\columnwidth]{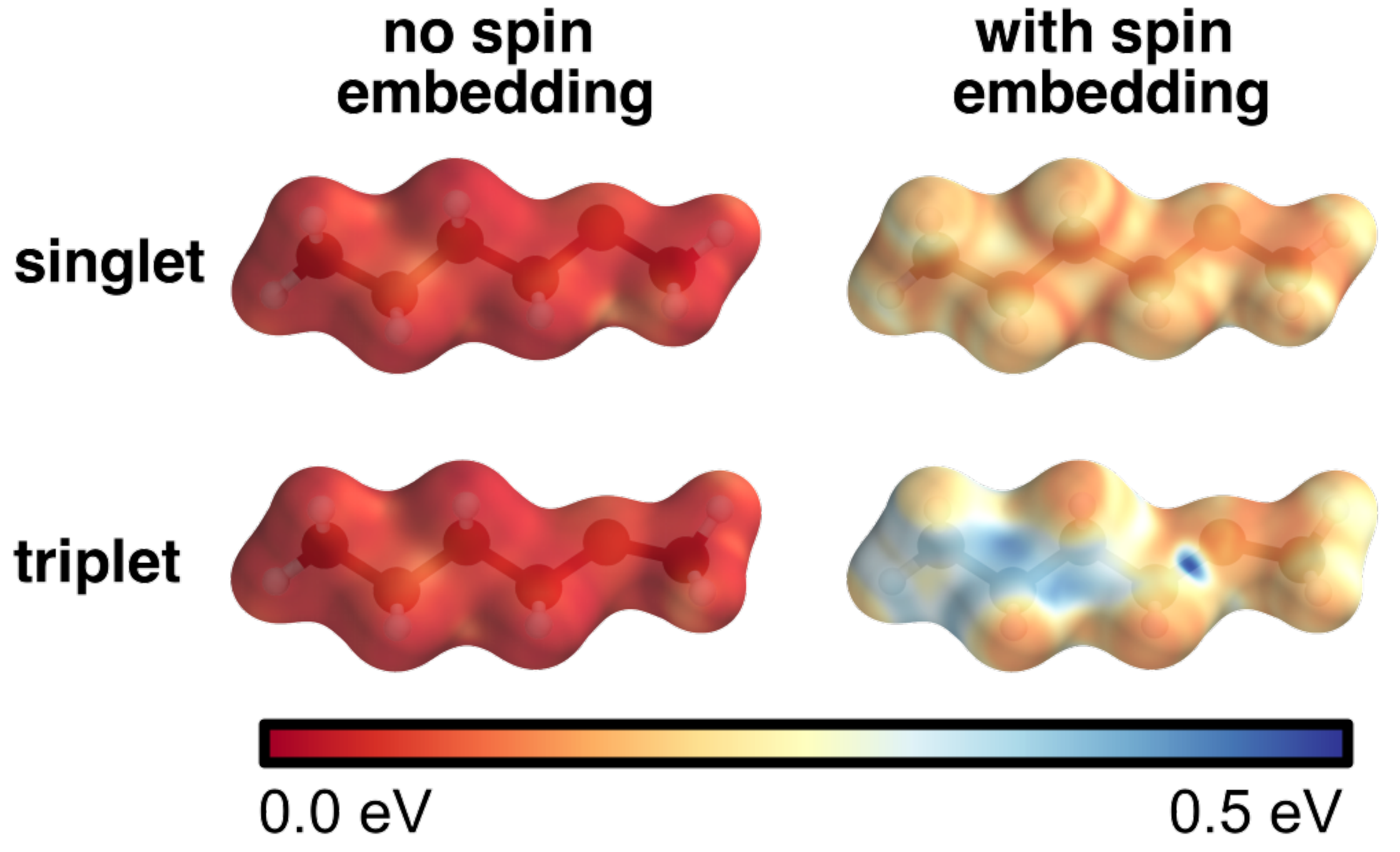}
	\caption{Local chemical potential for a random carbene chosen from the QMSpin database (the optimized geometries for the singlet/triplet states are shown). The chemical potential for a model without spin embeddings lacks features, whereas a model with spin embeddings learns a rich representation with significant differences between singlet and triplet states.}
	\label{fig:different_spin_gummybears}
\end{figure*}

\begin{figure*}
	\includegraphics[width=\columnwidth]{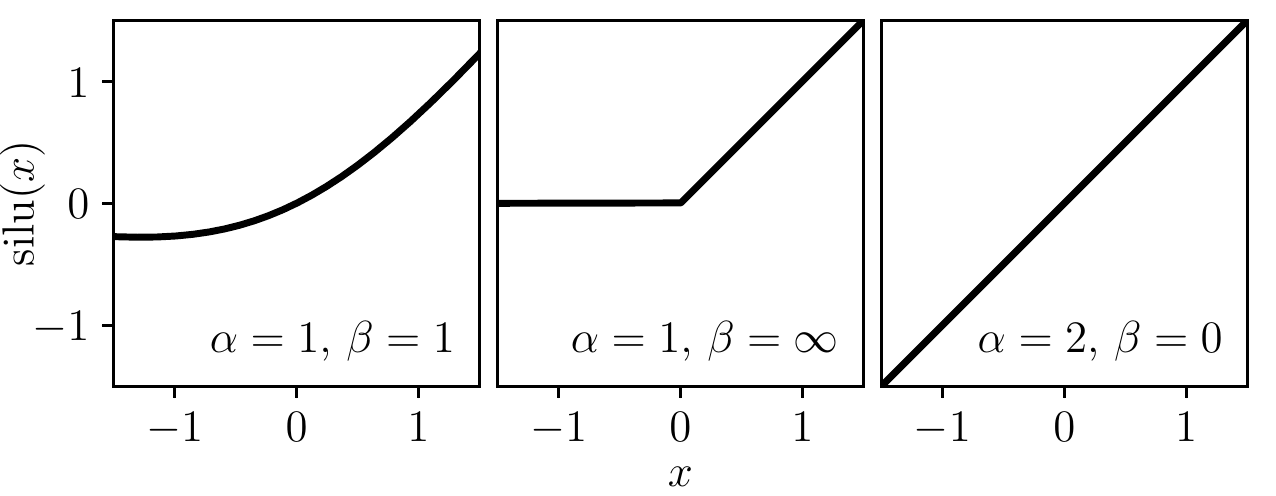}
	\caption{Generalized SiLU activation (Eq.~\ref{Meq:activation}). For $\alpha=1,\beta=\infty$, $\mathrm{silu}(x)$ is equivalent to $\mathrm{max}(x,0)$ (also known as $\mathrm{ReLU}$ activation), whereas for $\alpha=2,\beta=0$, the identity function is obtained.}
	\label{fig:activation}
\end{figure*}

\begin{table*}
	\begin{tabular}{c c c c c c c c c c c c c c c c c c c c c c c c}
		\toprule
		element & & & Z & 1s & 2s & 2p & 3s & 3p & 4s &  3d & 4p & 5s & 4d & 5p & 6s & 4f & 5d & 6p & vs & vp & vd & vf & \\
		\midrule
		H & $\mathbf{d}'_{1} =$ & $[$ & 1 & 1 & 0 & 0 & 0 & 0 & 0 &  0 & 0 & 0 &  0 & 0 & 0 &  0 &  0 & 0 & 1 & 0 &  0 &  0 & $]^{\mathsf{T}}$\\
		C & $\mathbf{d}'_{6} =$ & $[$ & 6 & 2 & 2 & 2 & 0 & 0 & 0 &  0 & 0 & 0 &  0 & 0 & 0 &  0 &  0 & 0 & 2 & 2 &  0 &  0 & $]^{\mathsf{T}}$\\
		N & $\mathbf{d}'_{7} =$ & $[$ & 7 & 2 & 2 & 3 & 0 & 0 & 0 &  0 & 0 & 0 &  0 & 0 & 0 &  0 &  0 & 0 & 2 & 3 &  0 &  0 & $]^{\mathsf{T}}$\\
		O & $\mathbf{d}'_{8} =$ & $[$ & 8 & 2 & 2 & 4 & 0 & 0 & 0 &  0 & 0 & 0 &  0 & 0 & 0 &  0 &  0 & 0 & 2 & 4 &  0 &  0 & $]^{\mathsf{T}}$\\
		P & $\mathbf{d}'_{15} =$ & $[$ & 15 & 2 & 2 & 6 & 2 & 3 & 0 &  0 & 0 & 0 &  0 & 0 & 0 &  0 &  0 & 0 & 2 & 3 &  0 &  0 & $]^{\mathsf{T}}$\\
		S & $\mathbf{d}'_{16} =$ & $[$ & 16 & 2 & 2 & 6 & 2 & 4 & 0 &  0 & 0 & 0 &  0 & 0 & 0 &  0 &  0 & 0 & 2 & 4 &  0 &  0 & $]^{\mathsf{T}}$\\
		Fe & $\mathbf{d}'_{26} =$ & $[$ & 26 & 2 & 2 & 6 & 2 & 6 & 2 &  6 & 0 & 0 &  0 & 0 & 0 &  0 &  0 & 0 & 2 & 0 &  6 &  0 & $]^{\mathsf{T}}$\\
		I & $\mathbf{d}'_{53} =$ & $[$ & 53 & 2 & 2 & 6 & 2 & 6 & 2 & 10 & 6 & 2 & 10 & 5 & 0 &  0 &  0 & 0 & 2 & 5 & 10 &  0 & $]^{\mathsf{T}}$\\
		Au & $\mathbf{d}'_{79} =$ & $[$ & 79 & 2 & 2 & 6 & 2 & 6 & 2 & 10 & 6 & 2 & 10 & 6 & 1 & 14 & 10 & 0 & 1 & 0 & 10 & 14 & $]^{\mathsf{T}}$\\
		Rn & $\mathbf{d}'_{86} =$ & $[$ & 86 & 2 & 2 & 6 & 2 & 6 & 2 & 10 & 6 & 2 & 10 & 6 & 2 & 14 & 10 & 6 & 2 & 6 & 10 & 14 & $]^{\mathsf{T}}$\\
		\bottomrule
	\end{tabular}
	\caption{Examples of element descriptors. Here, unscaled descriptors $\mathbf{d}'_Z$ are shown. The entries encode information about the ground state electron configuration (e.g.\ 1s$^2$2s$^2$2p$^2$ for C), the total number of electrons/nuclear charge (e.g.\ $Z=7$ for N), and the number of electrons in the valence shells (e.g.\ vs$^2$vp$^4$ for O). The descriptors used in Eq.~\ref{Meq:nuclear_embedding} are given by $\mathbf{d}_Z=\mathbf{d}'_Z\oslash\mathbf{d}'_{86}$, where $\oslash$ denotes Hadamard (element-wise) division (such that all entries of $\mathbf{d}_Z$ lie between $0$~and~$1$, which is desirable for numerical reasons). In this work, it is assumed that $Z_{\rm max}=86$ covers most practical applications, but descriptors for heavier elements could be derived analogously (and the scaling procedure adapted accordingly if necessary).}
	\label{tab:species_descriptor}
\end{table*}

\begin{table*}
	\begin{tabular}{c c c c c c c c c c}
		\toprule
		loss weights (see Eq.~\ref{Meq:loss}) & \qquad\qquad\qquad & \quad & \multicolumn{3}{c}{$\alpha_E = \alpha_F = \alpha_\mu = 1$} & \quad\quad & \multicolumn{3}{c}{$\alpha_E = \alpha_\mu = 1$, $\alpha_F = 100$} \\
		& & &
		\textbf{energy} & \textbf{forces} &
		\textbf{dipole} & & \textbf{energy} & \textbf{forces} &
		\textbf{dipole}\\
		\midrule
		\textbf{known molecules/} & MAE & & 18.513 & 39.863 & 39.380 &  & 10.620 & 14.851& 121.38 \\
		\textbf{unknown conformations} & RMSE & & 29.885 & 65.810  & 58.874 & & 16.782 & 25.330 & 165.82\\
		\midrule
		\textbf{unknown molecules/} & MAE & & 20.490 & 46.567 & 41.766 & & 13.151 & 17.326 & 120.50\\
		\textbf{unknown conformations} & RMSE & & 28.740 & 74.700 & 61.062 & & 17.891 & 26.179 & 162.32 \\
		\bottomrule
	\end{tabular}
	\caption{Mean absolute errors (MAEs) and root mean square errors (RMSEs) of energies (meV), forces
		(meV~\AA$^{-1}$) and dipole moments (mD) for the QM7-X\cite{hoja2021qm7} dataset. Here, \nn{} is trained with either a low or high force weight $\alpha_F$ in the loss function (see Eq.~\ref{Meq:loss}).}
	\label{tab:qm7x_results_low_force_weight}
\end{table*}
	
	\bibliography{references}